%% file: JointSensingandCom_shortenedversion.tex
\documentclass[journal,12pt,onecolumn,draftclsnofoot]{IEEEtran}

\usepackage{bbm}
\usepackage{amsmath}
\usepackage{acronym}  
\usepackage[dvips]{color}
\usepackage{epsf}
\usepackage{times}
\usepackage{epsfig}
\usepackage{notoccite}
\usepackage{graphicx}
\usepackage{epstopdf}
\usepackage{pstricks}
\usepackage{amssymb}
\usepackage{amsxtra}
\usepackage{here}
\usepackage{rawfonts}
\usepackage{float}
\usepackage{times}
\usepackage{url}
\usepackage{cite}
\usepackage{caption}
\usepackage{subcaption}
\usepackage{algorithm}
\usepackage{algpseudocode}
\usepackage{blindtext}
\usepackage{enumitem}
\usepackage{xcolor,cite,etoolbox}
\usepackage{relsize}
\usepackage{lipsum}
\usepackage{graphicx}
\usepackage{tabularx}
\usepackage{xparse}
\usepackage{array}
\newcolumntype{P}[1]{>{\centering\arraybackslash}p{#1}}
\usepackage{mleftright}

\usepackage{mathtools}

\usepackage{graphics}
\usepackage{physics}
\usepackage{amssymb}
\usepackage{siunitx}
\include{newcommands}
\usepackage{multicol}

\usepackage[nomain,acronym,shortcuts]{glossaries}
\makeglossaries
\newcommand*{\acro}[3][]{\newacronym[#1]{#2}{#2}{#3}}
\include{acronyms}

\usepackage{datetime}
\usepackage{amssymb}
\usepackage{subcaption}
\usepackage{verbatim}

\newtheorem{theorem}{\bf Theorem}

\newtheorem{proposition}{\bf Proposition}
\newtheorem{lemma}{\bf Lemma}

\newtheorem{definition}{\bf Definition}

\begin{document}
	\title{Joint Sensing, Communication, and AI: A Trifecta for Resilient THz User Experiences } 
	\author{\normalsize{Christina Chaccour, \emph {Member, IEEE,}
		Walid Saad,  \emph{Fellow, IEEE,}\\
	 Mérouane Debbah, \emph{Fellow, IEEE}, and H. Vincent Poor, \emph{Life Fellow, IEEE}\vspace{-1cm}}
	\thanks{C. Chaccour and W. Saad are with the Wireless@VT, Bradley Department of Electrical and Computer Engineering, Virginia Tech, Arlington, VA, USA, Emails: \protect{christinac@vt.edu}, \protect{walids@vt.edu}.}
	\thanks{ M. Debbah is with the Technology Innovation Institute, 9639 Masdar City, Abu Dhabi, United Arab Emirates and also with CentraleSupelec, University Paris-Saclay, 91192 Gif-sur-Yvette, France, Email: \protect{merouane.debbah@tii.ae}.}
	\thanks{H.V. Poor is with Department of Electrical and Computer Engineering, Princeton University, NJ 08544, USA, Email: \protect{poor@princeton.edu}.}}
	\maketitle
\normalsize
\vspace{-1cm}
	\begin{abstract} 
		In this paper a novel joint sensing, communication, and artificial intelligence (AI) framework is proposed so as to optimize  \ac{XR} experiences over \ac{THz} wireless systems. The proposed framework consists of three main components. \emph{First}, a tensor decomposition framework is proposed to extract unique sensing parameters for XR users and their environment by exploiting then \ac{THz} channel sparsity. Essentially, \ac{THz} band's quasi-opticality is exploited and the sensing parameters are extracted from the uplink communication signal, thereby allowing for the use of the \emph{same waveform, spectrum, and hardware for both communication and sensing functionalities}. Then, the Cramer-Rao lower bound is derived to assess the accuracy of the estimated sensing parameters. \emph{Second}, a non-autoregressive multi-resolution generative \ac{AI} framework integrated with an adversarial transformer is proposed to predict missing and future sensing information. The proposed framework offers robust and comprehensive historical sensing information and anticipatory forecasts of future environmental changes, which are \emph{generalizable to fluctuations in both known and unforeseen user behaviors and environmental conditions}. \emph{Third}, a multi-agent deep recurrent hysteretic Q-neural network is developed to control the handover policy of \ac{RIS} subarrays, leveraging the informative nature of sensing information to minimize handover cost, maximize the individual \acp{QoPE}, and improve the robustness and resilience of \ac{THz} links. Simulation results show a high generalizability of the proposed unsupervised generative \ac{AI} framework to fluctuations in user behavior and velocity, leading to a $\SI{61}{\%}$ improvement in instantaneous reliability compared to schemes with known channel state information.
	\end{abstract}
	\IEEEpeerreviewmaketitle
\section{Introduction}\label{sec:Intro}
\vspace{-.2cm}
The sixth generation (6G) wireless system needs to support advanced services such as \acrfull{XR}\footnote{\ac{XR} encompasses \ac{AR}, \ac{MR}, and \ac{VR}.}, autonomous driving, and digital twins~\cite{saad2019vision}. Catering to the demands of such applications requires wireless systems to not only provide communication services, but to also include sensing, localization, and control capabilities. In particular, designing a versatile wireless XR system that delivers a multi-sensory immersive experience poses numerous challenges, including the need to maximize data rates, ensure instant reliability, minimize latency, accurately track the \ac{6DoF} of each \ac{XR} user, and maintain situational awareness of the surrounding environment. One approach to overcoming these challenges involves the utilization of higher frequency bands, specifically the sub-\acrfull{THz} and \ac{THz} bands ($0.1-\SI{10}{THz}$), as they can provide significantly elevated data rates and high-resolution environmental sensing capabilities\cite{sarieddeen2021overview}, which can culminate in a personalized and interactive experience. However, these bands, on their own, cannot fully resolve all the design challenges needed for a multi-sensory XR system experience, such as achieving instantaneous reliability, near-zero latency, and precise tracking of the \ac{6DoF} of each users' head and body.\\
\indent \ac{THz} networks could exploit the sensing information for two distinct key purposes: instilling sensing-driven \emph{service} intelligence and sensing-driven \emph{operational} intelligence. \emph{First,} service intelligence refers to the ability of the \ac{XR} system to accurately adapt and respond to the user's needs and preferences in real-time, based on the information gathered from each user's environment and interaction. For example, in VR, precise tracking of the user's movements and interactions is necessary to avoid motion sickness and other discomfort. In contrast, \ac{AR} requires accurate tracking to maintain alignment between the real-world and the augmented environment. \emph{On the other hand,} operational intelligence refers to the network's ability to sense the dynamic environment and incorporate its behavioral stochasticity in real-time. For instance, the narrow \ac{THz} beams are particularly susceptible to minute changes induced by the user behavior and/or environment, necessitating frequent beam training and channel estimation. However, this approach results in significant overhead and delays, which are not acceptable in wireless \ac{XR} whose applications require \emph{instantaneously continual high-rate \ac{THz} beams} that guarantee seamless alignment between virtual and real environments. By leveraging sensing-driven operational intelligence, a \ac{THz} network can effectively mitigate its intermittent link behavior and achieve a tight alignment between virtual and real environments, resulting in a seamless user experience. Additionally, by sharing the same waveform, hardware, and spectrum for communication and sensing functionalities, the system can form more robust beams in a time-critical manner, boosting spectral efficiency, and enhancing the user experience.\\
\indent Fundamentally, the main goal of effective joint sensing and communication systems is to \emph{mitigate every communication challenge and leverage it as a sensing opportunity.} Achieving this goal requires overcoming key challenges. \emph{Firstly}, sensing and communication are functionally different\footnote{For instance, sensing typically relies on unmodulated signals or short pulses and chirps. Communication signals, on the other hand, are a mix of unmodulated (pilots) and modulated signals\cite{chaccour2021seven}.}. Thus, there is a need for a holistic approach that can extract sensing parameters and analyze situational awareness from a wireless system designed to leverage the same waveform, spectrum, and hardware for both sensing and communication functions. \emph{Secondly}, optimizing \ac{XR} applications served by a \ac{THz} network presents distinct challenges compared to traditional network optimization frameworks. Essentially, the inherently unpredictable nature of the \ac{THz} channel requires \emph{continual and complete sensing} input to ensure a resilient and seamless \ac{THz} experience for \ac{XR} users. Thus, it is necessary to leverage the sensing input for prediction and forecasting purposes so as to enable a \emph{proactive} decision-making and adaptation to environmental dynamics, and ensure a resilient THz experience for XR users. \emph{Thirdly}, putting the user and their \acrfull{QoPE} at the center is crucial to delivering personalized and immersive experiences in \ac{XR}. This requires designing optimization frameworks that not only account for the diverse needs and preferences of different users but also adapt to the stochastic nature of the THz environment. Overall, these challenges require innovative solutions that can effectively address the resilience of \ac{THz} links while exploiting the trifecta of sensing, communication, and \ac{AI} faculties, and ultimately ensuring near-optimal immersive wireless \ac{XR} experience.
\vspace{-0.25cm}
\subsection{Prior Works}
The concept of joint communication and sensing has recently seen a surge of interest in the literature \cite{bazzi2023robust, zuo2022reconfigurable, guerra2021real, quekthzletter, chen2022isac}. The work in~\cite{bazzi2023robust} proposed a beamforming design dedicated for dual-functional radar and communication base stations. In~\cite{zuo2022reconfigurable}, the authors exploited the potential of \acrfullpl{RIS} to improve radar sensing in a non-orthogonal multiple access empowered integrated sensing and communication network.  However, the works in \cite{bazzi2023robust} and \cite{zuo2022reconfigurable} do not leverage the \ac{THz} band's quasi-opticality nor attempt to effectively mitigate the highly uncertain nature of the \ac{THz} channel by capitalizing on the sensing input. Moreover, this prior art does not exploit the sensing functionality to improve the unique network performance indicators of wireless \ac{XR} services. When it comes to \ac{THz} sensing,  the work in  \cite{guerra2021real} considered the joint detection, mapping, and navigation problem by a drone with real-time learning capabilities. In \cite{quekthzletter}, model-based and model-free hybrid beamforming techniques are proposed for a joint radar and communication system. The authors in~\cite{chen2022isac} developed a joint reference signal and synchronization signal block-based sensing scheme to predict the need for beam switches. Nonetheless, these works~\cite{bazzi2023robust, zuo2022reconfigurable, guerra2021real, quekthzletter, chen2022isac} do not opportunistically use the same spectrum, waveform, and hardware at \ac{THz} bands for joint sensing and communications. Consequently, these existing deployment strategies lead to higher costs in terms of hardware and resources, which is not suitable for commercial \ac{XR} services. In~\cite{chaccour2021joint}, we proposed a novel joint sensing and communication system that utilizes the same waveform, hardware, and spectrum. However, our work in~\cite{chaccour2021joint} does not have the \ac{AI} capabilities needed to predict missing and future sensing information. In fact, our prior work mainly relies on the sensing input in its raw unoptimal structure. In particular, and more importantly, the works in ~\cite{bazzi2023robust, zuo2022reconfigurable, guerra2021real, quekthzletter, chen2022isac,chaccour2021joint} do not design robust and resilient \ac{AI}-oriented frameworks that can optimize the \emph{instantaneous} user experience of future user-centric and intelligence-centric 6G applications. Clearly, the current body of literature on joint sensing and communication systems at THz frequencies lacks a comprehensive and foundational approach to adequately address the challenge of managing uncertainty in \ac{THz} channels through the integration of AI and sensing.
\begin{figure}[!t]
	\centering
	\includegraphics[width=0.72\textwidth]{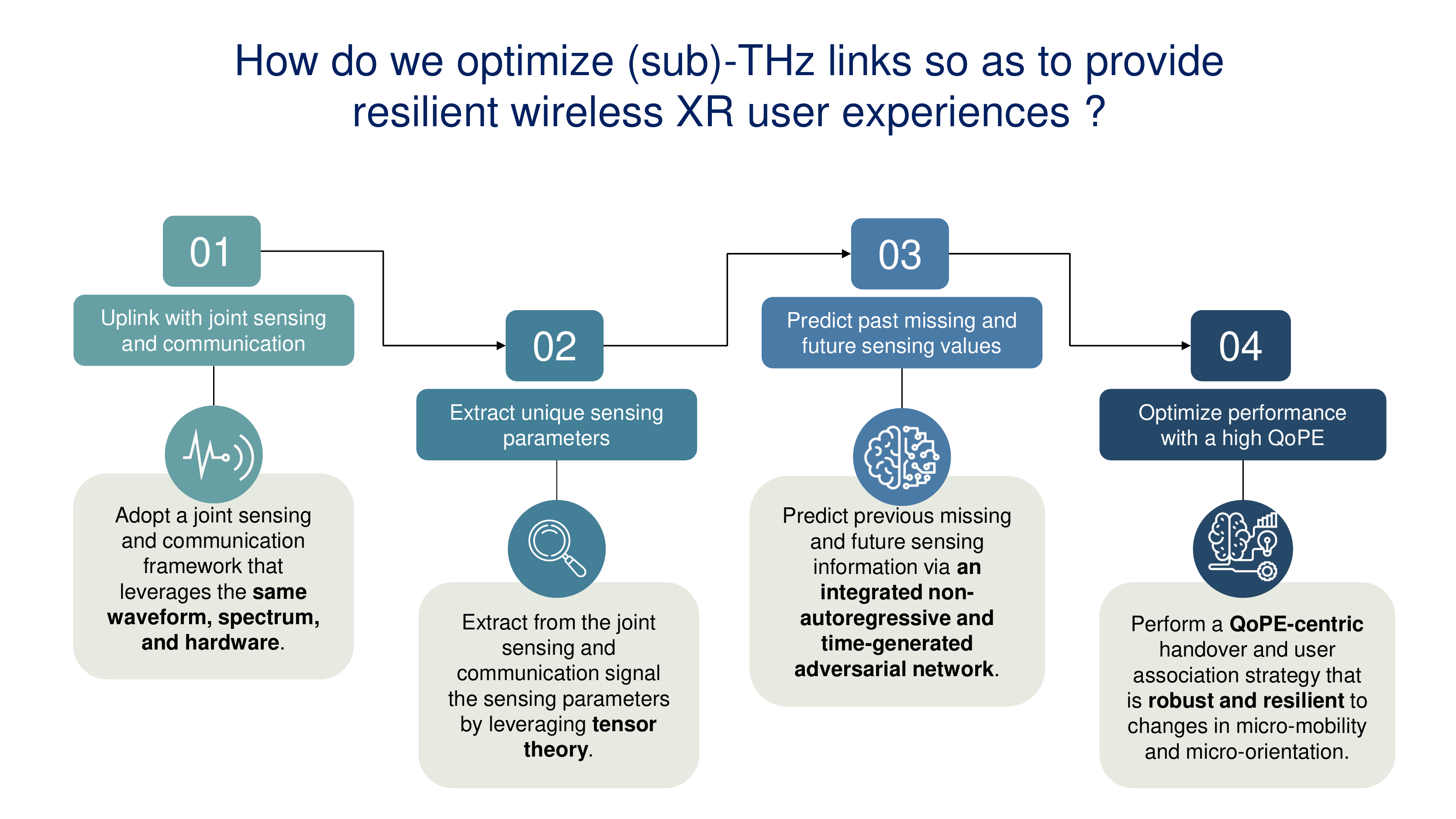}
	\caption{\small{Illustrative figure showcasing the key steps to optimize the \ac{QoPE} of \ac{XR} users served with \ac{THz} links.}}
	\label{fig:contributions}
	\vspace{-0.5cm}
\end{figure} 
\subsection{Contributions}
\indent The main contribution of this paper is, thus, a novel design, analysis, and optimization framework that integrates communication, sensing, and \ac{AI}, to address the fundamental question of whether or not such integrated systems can provide highly reliable \ac{THz} links instantaneously. If this is not the case, \emph{can such systems still deliver robust and resilient \ac{THz} experiences to \ac{XR} users?} Fig.~\ref{fig:contributions} illustrates the roadmap of our approach, with our key contributions being:
\begin{itemize}
	\item To extract the environmental \emph{sensing} parameters that characterize the DoF and associated blockages of wireless \ac{XR} users, we leverage the \ac{THz} uplink \emph{communication} waveform via tensor theory. In particular, we improve the spectral and energy efficiency of our design by opportunistically utilizing the same waveform, hardware, and spectrum for sensing and communication. We also leverage the inherent sparsity in the \ac{THz} bands to perform a tensor decomposition of the received uplink signal, and we prove the uniqueness of our extracted sensing parameters.
	\item To assess the accuracy of our estimated sensing parameters, we derive their \ac{CRLB} and analyze the parameters that may affect the variability of the estimation. Our derived results show that the estimation error varies non-monotonically with the carrier frequencies, i.e., frequencies above $\SI{1}{THz}$ exhibit higher molecular absorption lines. This underscores the significance of operating within the sub-THz spectrum, as it achieves a favorable equilibrium between the inherent tradeoffs of sensing and communication functionalities. 
	\item  To address the issue of missing tracking and blockage sensing information caused by the intermittent nature of links, we propose a novel \ac{AI}-based imputation design that exploits a novel non-autoregressive multi-resolution generative framework to predict the missing values in a sequence of sensing values. We then propose a novel adversarial transformer framework that processes comprehensive sensing information and then, predicts, future user behavior and dynamic changes to the environment (e.g. potential blockages). These designs enable the recovery of missing tracking and blockage sensing information thus allowing the network to anticipate user behavior, ensuring \emph{seamless, continual, and uninterrupted \ac{XR} experiences}.
	\item We then introduce a novel \ac{QoPE}-centric-aware optimization problem that leverages the designed sensing and communication capabilities to optimize handover-- a challenging task in a highly uncertain, dynamic network that relies narrow beamwidths. This approach accounts for both current and anticipated user behavior, and allows minimization of the frequency of unnecessary handovers thereby ensuring the robustness and resilience of \ac{THz} links. To address this problem, we introduce a semi-distributed multi-agent \ac{RL} framework that substantially improves the resilience of individualized user experiences across the entire \ac{XR} spectrum through its tailored reward system.
	\item  Simulation results demonstrate that the proposed integrated imputation-forecasting system has higher generalizability and maintains accuracy despite fluctuations in user speed and number of users entering/leaving the room, while vanilla deep learning frameworks quickly lose accuracy. The QoPE-centric optimization framework significantly improves user reliability and resilience, with gains of up to $\SI{78}{\%}$ compared to baseline schemes with known \ac{CSI} and beam tracking.
\end{itemize}
\section{System Model}\label{Sec:Sys-Model}
Consider the downlink and uplink of an \ac{RIS}-based single cell-\ac{THz} system operating as a joint sensing and communication system in a confined indoor area. A set $\mathcal{B}$ of $B$ \acp{RIS} are used as \ac{THz} \acp{BS} that transmit \ac{XR} content and provide situational awareness for a set $\mathcal{U}$ of $U$ mobile wireless \ac{XR} users. Each subarray $n$ of \ac{RIS} $b$ in $\mathcal{B}$ is located at $\boldsymbol{l}^{b,n}=\left[l^{b,n}_x, l^{b,n}_y\right]^T \in \mathbb{R}^2$. The \ac{XR} users are mobile and may change their locations and orientations at any point in time. As such, $\boldsymbol{l}^u=\left[l^u_x, l^u_y\right] ^T \in \mathbb{R}^2$ denotes the position of user $u$. Here, \emph{situational awareness} refers to the process of mapping the physical world, including the location, orientation, and state of physical objects, to the digital realm with a certain level of accuracy. We consider the \emph{entire \ac{XR} reality-virtuality continuum}, as will be evident in our later analysis. \acp{RIS} are used to provide nearly continuous \acrfull{LoS} data links to the \ac{XR} users. This \ac{RIS}-enhanced architecture also creates multiple independent paths capable of gathering rich information about the environment \cite{chaccour2021seven}. 
\begin{figure}[!t]
\centering
\includegraphics[width=0.55\textwidth]{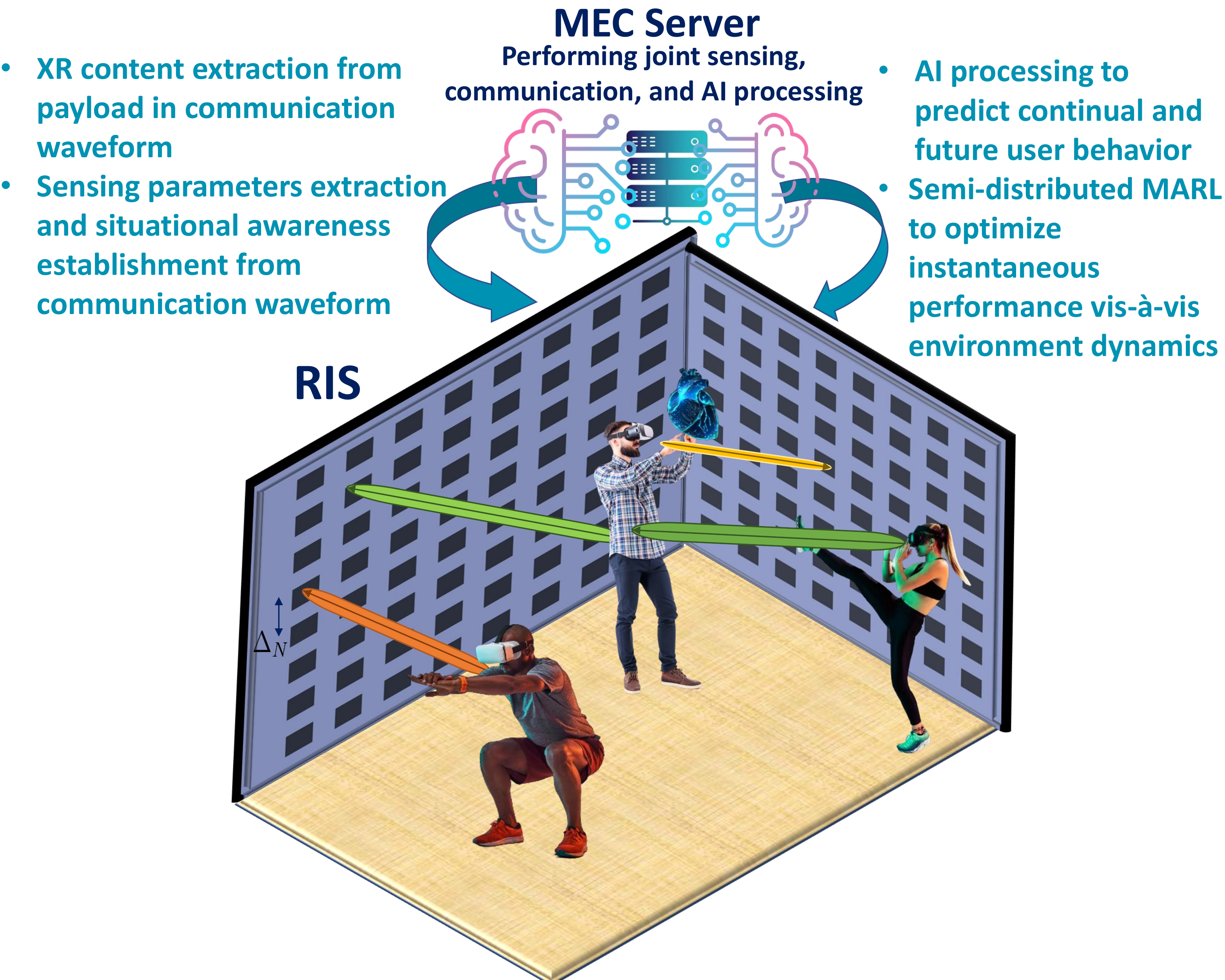}
\caption{\small{Illustrative example of our THz-based wireless XR system for joint communication, sensing, and AI.}}
\label{fig:model}
\vspace{-0.5cm}
\end{figure} 
Each \ac{RIS} consists of a two-dimensional large antenna array that has an \ac{AoSA} structure~\cite{faisal2020ultramassive}. Antenna elements of the same subarray are connected to the same phase shifter and \ac{RF} chain, thus reducing the number of phase shifters needed compared to a fully-connected array structure. Each subarray has a large number of antennas $M$ and the spacing between antenna elements within each \ac{RIS} subarray is $\delta_m$. The total number of subarrays is $N$ with the spacing between them being $\Delta_N$. These subarrays collectively employ a hybrid beamforming architecture.  Without loss of generality, we assume $N \geq U$, thus providing \ac{XR} users with sufficient DoF in terms of association. This subarray structure allows us to have $N$ independent channels each \ac{RIS}.  
\subsection{Joint Communication and Sensing Model}
Consider an arbitrary \ac{RIS} in $\mathcal{B}$ that captures the situational awareness and mapping information, and receives high data rate \ac{XR} content as in Fig.~\ref{fig:model}. We assume that the \ac{RIS} has performed synchronization and initial beam alignment according to existing techniques (e.g. \cite{boulogeorgos2019performance}). To maintain high-resolution real-time user tracking and a situational awareness of the indoor area, each \ac{RIS} will continuously collect a sequence of $T$ snapshots of the received uplink communication signal that are then used to estimate and track the \ac{AoA}, \ac{AoD}, and \ac{ToA} of users as well as the location of obstacles leading to \ac{NLoS} communications. Since we use indoor \acp{RIS},  blockages mostly occur due to bodies of the users.\\
\indent Each \ac{XR} user has a \ac{ULA} of $Q$ antennas and $R$ \ac{RF} chains. Hence, if all \ac{XR} users are active, $U$ data signaling beams are sent simultaneously to the \ac{RIS}. To mitigate the high \ac{PAPR} effect of \ac{OFDM} schemes, a \ac{SC-FDM} is used to maintain high energy efficiency at the user equipment. The far-field \ac{EM} wave condition is assumed to be met. After beamforming, the transmitted user signal at subcarrier $k$ is $\boldsymbol{s}_{k}(t)=\boldsymbol{F}_{\mathrm{R}}(t) \boldsymbol{F}_{k}(t) \boldsymbol{x}_{k}(t), \forall k=1, \ldots, K$, where $\boldsymbol{x}_{k}(t) \in \mathbb{C}^{l\times1}$ is the transmitted uplink communication signal, $\boldsymbol{F}_{k}\in \mathbb{C}^{R\times l}$ is the digital precoding matrix for subcarrier $k$, and $\boldsymbol{F}_{\mathrm{R}}(t)\in \mathbb{C}^{Q\times R}$ is the \ac{RF} precoder for all subcarriers. After processing the communication waveform of subcarrier $k$, the uplink received signal at each \ac{RIS} subarray $n$ will be $ y_{n,k}(t)=\boldsymbol{w}_{n,k}^{T} \boldsymbol{H}_{n,u}^k(t) \boldsymbol{s}_{k}(t)+\nu_{n, k}(t),$ where $\boldsymbol{w}_{n,k} \in \mathbb{C}^{M\times 1}$ is the combining vector of \ac{RIS} subarray $n$ at subcarrier $k$, $\boldsymbol{H}_{n,u}^k \in \mathbb{C}^{M \times Q}$ is the uplink channel matrix for subcarrier $k$, and $\nu_{n,k} \sim \mathcal{C}\mathcal{N}\left(\boldsymbol{0}, \boldsymbol{I}_{M\times Q}\right)$ is the additive Gaussian noise vector at subarray $n$ and subcarrier $k$.
To construct a single realization  for sensing the wireless environment, $T$ snapshots are recorded by the \ac{RIS}, each of which captures $J$ uplink measurements:
\begin{equation}\label{RSS}
\boldsymbol{y}_{n,k}(t)=\boldsymbol{W}_{n,k}^{T} \boldsymbol{H}_{n,u}^k(t) \boldsymbol{F}_{\mathrm{R}}^{m}(t) \boldsymbol{F}_{k}(t) \boldsymbol{s_k}(t)+\boldsymbol{\nu}_{n,k}(t),
\end{equation} 
where $\boldsymbol{y}_{n,k}(t)  \triangleq\left[
y_{n,k, 1}(t)  \ldots  y_{n,k, J}(t)\right]^{T}, \boldsymbol{\nu}_{n,k}(t)  \triangleq\left[\nu_{n,k, 1}(t)  \ldots \nu_{n,k, J}(t)\right]^{T}, 
\boldsymbol{W_{n,k}}  \triangleq\left[
\boldsymbol{w}_{n,k,1}  \ldots  \boldsymbol{w}_{n,k,J}\right].$
The $M \times Q$ \ac{MIMO} channel matrix between user $u$ and \ac{RIS} subarray $n$ in the time domain is: $
\boldsymbol{H}_{n,u}(t)=\sum_{p=1}^{P} \alpha^p_{n,u} \boldsymbol{a_r}\left(\phi^p_{n,u}\right) \boldsymbol{a_t}^H\left(\theta^p_{n,u}\right)\delta(t-\tau_{n,u}), $
where $\left(.\right)^H$ is the conjugate transpose, $\alpha^p_{n,u} \in \mathbb{C}$ is the complex channel gain, $P$ is the number of distinct spatial paths,  $\theta_{n,u}^{l}$ is the \ac{AoD} of path $p$ of antenna $u$, $\tau_{n,u}$ is the \ac{ToA}, and $\phi_{n,u}^{p}$ is the \ac{AoA} of path $p$ at subarray $n$. The array steering vector of the antenna array of the \ac{XR} user and the \ac{RIS} subarray, have a \ac{ULA} structure\footnote{While uniform planar arrays can be used fo our 2D subarrays, we adopt \ac{ULA} for analytical tractability.}, and are given by: $	\boldsymbol{a_t}\left(\theta^{p}_{n,u}\right)=\frac{1}{\sqrt{Q}}\left[1, e^{j \frac{2\pi}{\delta_m} \sin \left(\theta^p_{n,u}\right)}, \cdots, e^{j(Q-1) \frac{2\pi}{\delta_m} \sin \left(\theta^p_{n,u}\right)}\right]^{T}$,  $\boldsymbol{a_r}\left(\phi^{p}_{n,u}\right)=\frac{1}{\sqrt{M}}\left[1, e^{j \frac{2\pi}{\delta_m} \sin \left(\phi^p_{n,u}\right)}, \cdots, e^{j(M-1) \frac{2\pi}{\delta_m} \sin \left(\phi^p_{n,u}\right)}\right]^{T}$. Consequently, the channel matrix in the frequency domain associated with subcarrier $k$ will be:\vspace{-0.25cm}
\begin{equation}\label{channel}
\boldsymbol{H}^k_{n,u}(f)=\sum_{p=1}^{P} \alpha^p_{n,u} \boldsymbol{a_r}\left(\phi^p_{n,u}\right) \boldsymbol{a_t}^H\left(\theta^p_{n,u}\right)\exp\left(-\frac{j2\pi f\tau_{n,u}}{K} \right).
\end{equation}
\indent The snapshots in \eqref{RSS} can be collected over \ac{LoS} or \ac{NLoS} communication signal waveforms. For the \ac{LoS}, the uplink channel gain will be \cite{chaccour2020can}: $ \alpha^L_{n,u}=\frac{c}{4\pi f r_{n,u}} e^{-\frac{k(f)r_{n,u}}{2}} e^{-j2\pi f \tau^L_{n,u}}.$
Here, $c$ is the speed of light, $k(f)$ is the overall molecular absorption coefficients of the medium at \ac{THz} band, $f$ is the operating frequency, and $r_{n,u}$ is the distance between \ac{XR} user $u$  and \ac{RIS} subarray $n$. From this channel gain, we can see that, for \ac{LoS} conditions, the parameters to be estimated are $\{\theta^L_{n,u}, \phi^L_{n,u}, \tau^L_{n,u}\}$. This process allows tracking \ac{XR} users in real-time. Here, we do not consider \ac{NLoS} links for communications purposes due to the poor propagation characteristics of \ac{THz} signals (e.g., limited reflection) and their high susceptibility to blockage. In other words, the \ac{NLoS} link cannot meet the high-rate \ac{XR} needs. Nonetheless, from a sensing standpoint, instead of discarding the \ac{NLoS} signal completely, we use it to track the user and gather a situational awareness of the blockage incidence. In \ac{NLoS} conditions, the channel gain is given by \cite{moldovan2014and}: $\alpha^N_{n,u}=\frac{c}{4\pi f (r^{(1)}_{n,u}+r^{(2)}_{n,u})} e^{\left({-\frac{k(f)(r^{(1)}_{n,u}+r^{(2)}_{n,})}{2}}\right)} R(f)e^{-j2\pi f \tau^N_{n,u}},$
where  $r^{(1)}_{n,u}$ is the distance between the \ac{XR} user $u$ and the reflecting point,  and $r^{(2)}_{n,u}$ is the distance between the reflecting point and \ac{RIS} subarray $n$. For reflections, we consider the transverse electric part of the \ac{EM} wave, i.e., the signal is assumed to be perpendicular to the plane of incidence. This assumption is valid given the placement of our \ac{RIS} per Fig.~\ref{fig:model}. Thus, $R_{n,u}(f)=\gamma_{n,u}(f)\rho_{n,u}(f),$ is the reflection coefficient, where $\gamma(f)\approx-\exp \left(\frac{-2 \cos \left(\psi_{n,u}\right)}{\sqrt{\eta(f)^{2}-1}}\right)$ is the Fresnel reflection coefficient and $\rho_{n,u}(f)=\exp \left(-\frac{8 \pi^{2} f^{2} \sigma^{2}  \cos ^{2}\left(\psi_{n,u}\right)}{c^{2}}\right)$ is the Rayleigh factor that characterizes the roughness effect. $\psi_{n,u}$ is the angle of the incident signal to the reflector, $\eta(f)$ is the refractive index, and $\sigma$ is the surface height standard deviation. Thus, for \ac{NLoS}, the parameters to be estimated are: $\{\theta^N_{n,u}, \phi^N_{n,u}, \tau^N_{n,u}, \psi_{n,u}\}$. Next, we formulate the problem of estimating the sensing parameters from the $T$ collected snapshots in \eqref{RSS}. To do so, we leverage the sparsity of the \ac{THz} channel matrix and reformulate the received uplink signal expression as a tensor. \vspace{-0.15cm}
\section{Situational Awareness Estimation via Tensor Decomposition}\label{section:tensor}
\subsection{Problem Formulation}  
Estimating the sensing parameters from $T$ communication snapshots recorded at the \ac{RIS} subarray faces multiple challenges. First, existing estimation techniques \cite{mendrzik2019enabling, ozkaptan2018ofdm, guerra2021near} rely on the collection of pilot signals and their goal is to estimate the channel. In contrast, our goal here is to collect uplink communication signals used by users during an \ac{XR} session so as to \emph{continuously track} the sensing parameters.\footnote{Communication was initiated after a successful initial access and beam alignment prior to the \ac{XR} session.} Second, existing parameter estimation methods~\cite{zhu2017hybrid,gruber1997statistical, abdallah2022ris} have their own limitations: Off-grid methods like \ac{MUSIC} cannot jointly estimate multiple sensing parameters, leading to a difficult pairing problem \cite{gruber1997statistical}. Also, the \ac{MUSIC} algorithm cannot be readily applied to the peculiar \ac{AoSA} structure of hybrid \ac{THz} beamforming as its complexity of the method becomes impractical with large antenna arrays \cite{zhu2017hybrid}. Moreover, on-grid methods like compressive sensing~\cite{abdallah2022ris} are constrained by their grid spacing, which increases complexity if a high-resolution is desired.\\
\indent To successfully estimate the \ac{THz} sensing parameters, we concatenate $T$ snapshots of the received uplink signal:
\vspace{-0.5cm}
\begin{equation}\label{FreqRSS}
\boldsymbol{Y}_{n,k}=\boldsymbol{W}_{n,k}^{T} \boldsymbol{H}_{n,u}^k(f) \boldsymbol{\Omega}+\boldsymbol{N}_{n,k},
\end{equation} 
\vspace{-0.75cm}\small
\begin{align*}
&\boldsymbol{Y}_{n,k}  \triangleq\left[\begin{array}{lll}
\boldsymbol{y}_{n,k}(1)  \ldots & \boldsymbol{y}_{n,k}(T)
\end{array}\right], \boldsymbol{N}_{n,k}  \triangleq\left[\begin{array}{lll}
\boldsymbol{\nu}_{n,k}(1) & \ldots & \boldsymbol{\nu}_{n,k}(T)
\end{array}\right],\\
&\boldsymbol{\Omega}  \triangleq\left[\begin{array}{lll}
\boldsymbol{F}_{\mathrm{R}}^{m}(1) \boldsymbol{F}_{k}(1) \boldsymbol{s_k}(1) & \ldots & \boldsymbol{F}_{\mathrm{R}}^{m}(T) \boldsymbol{F}_{k}(T) \boldsymbol{s_k}(T)
\end{array}\right]. 
\end{align*}
\small
\normalsize
From \eqref{FreqRSS}, we observe that the received signal has three modes that represent the number of measurements, the number of snapshots, and the number of subcarriers, respectively. Thus, we can model it as a three-order tensor, namely, $\boldsymbol{\chi} \in \mathbb{C}^{J \times T \times K}$. In fact, substituting \eqref{channel} in \eqref{FreqRSS} yields:
\begin{align}\label{New_RSS}
\boldsymbol{Y}_{n,k}(t)&=\sum_{p=1}^{P} \Lambda_{n,u,k}^p\boldsymbol{\zeta}(\phi^p_{n,u})\boldsymbol{\xi}(\theta^p_{n,u})^H+ \boldsymbol{N}_{n,k}.
\end{align} 
\noindent In \eqref{New_RSS}, $\Lambda_{n,u,k}^p\triangleq\alpha^p_{n,u}\exp(-\frac{j2\pi f\tau_{n,u}}{K})$, $ \boldsymbol{\zeta_k}(\phi^p_{n,u})\triangleq\boldsymbol{W}_{n,k}^{T} \boldsymbol{a_r}\left(\phi^p_{n,u}\right)$, and $\boldsymbol{\xi_k}(\theta^p_{n,u})\triangleq\boldsymbol{\Omega}^H\boldsymbol{a_t}\left(\theta^p_{n,u}\right)$. Clearly, each slice $\boldsymbol{Y}_{n,k}$ corresponding with tensor $\boldsymbol{\chi} $ is a weighted sum of a common set of rank-one outer products. Hence, we can factorize the tensor as follows:
\begin{equation*}
\begin{aligned}
\boldsymbol{\chi} =\sum_{p=1}^{P} \boldsymbol{\zeta_k}(\phi^p_{n,u})\circ \boldsymbol{\xi_k}(\theta^p_{n,u})\circ \Lambda_{n,u}^p+\mathcal{W} =\left[ \left[ \mathbf{A}, \mathbf{B}, \mathbf{C} \right] \right] +\mathcal{W},
\end{aligned}
\end{equation*}
\small
\begin{equation}
\begin{aligned}
\boldsymbol{A}  \triangleq\left[\begin{array}{lll}\boldsymbol{\zeta}_1(\phi_{n,u}^P) \ldots &\boldsymbol{\zeta}_K(\phi_{n,u}^P)\end{array}\right], 
\boldsymbol{B}  \triangleq \left[\begin{array}{lll} \boldsymbol{\xi}_1(\theta^p_{n,u}) \ldots \boldsymbol{\xi}_K(\theta^p_{n,u})&\end{array}\right],
\boldsymbol{C}  \triangleq\left[\begin{array}{lll}\boldsymbol{\Lambda}_{n,u,1}^p&\ldots \boldsymbol{\Lambda}_{n,u,K}^p \end{array}\right].
\end{aligned}
\end{equation}
\normalsize 
Here, $\circ$ is the outer product symbol, $\boldsymbol{\Lambda}_{n,u,k}^p=\left\{\alpha^p_{n,u}\exp(-\frac{j2\pi f\tau_{n,u}}{K}) \right\}_{t=1}^{T}$, and $\mathcal{W}$ is $\boldsymbol{N}_{n,k}$ in the tensor domain.
$\boldsymbol{A}, \boldsymbol{B},$ and $\boldsymbol{C}$ are the three factor matrices associated to the tensor $\boldsymbol{\chi}$. 
We note that, the sparsity of the \ac{THz} channel and the limited number of propagation paths, leveraging tensor theory is particularly efficient and can guarantee the uniqueness of the estimated sensing parameters. Moreover, our approach can characterize the sensing parameters from the collected snapshots without imposing additional constraints or assumptions. For instance, alternatively adopting matrix factorization almost never leads to a unique solution unless the rank of the matrix is one or further conditions are imposed on the factor matrices.
\vspace{-0.25cm}
\subsection{Proposed Sensing Parameters Estimation Method}
After factorizing the tensor $\boldsymbol{\chi}$, extracting the sensing parameters and estimating them requires solving the following optimization problem:\vspace{-0.25cm}
\small
\begin{equation}\label{optimization}
\min _{\widetilde{\boldsymbol{A}}, \widetilde{\boldsymbol{B}}, \widetilde{\boldsymbol{C}}}\left\|\boldsymbol{\chi}-\sum_{p=1}^{P} \widetilde{\boldsymbol{a}}_{p} \circ \tilde{\boldsymbol{b}}_{p} \circ \widetilde{\boldsymbol{c}}_{p}\right\|_{F}^{2},
\end{equation}
\normalsize
where $\widetilde{\mathbf{A}}=\left[\widetilde{\boldsymbol{a}_{1}} \cdots \widetilde{\boldsymbol{a}}_{P}\right], \widetilde{\boldsymbol{B}}=\left[\widetilde{\boldsymbol{b}}_{1} \cdots \widetilde{\boldsymbol{b}}_{P}\right], \widetilde{\boldsymbol{C}}=\left[\widetilde{\boldsymbol{c}}_{1} \cdots \widetilde{\boldsymbol{c}}_{P}\right]$ are the three estimated factor matrices. To solve this problem, the three factor matrices need to be estimated. To do so, we leverage the sparsity of the \ac{THz} channel that guarantees the uniqueness condition for tensor decomposition. This allows us to establish a relationship between the true factor matrices and their estimates. Thus, given these relationships, we next derive the environmental sensing parameters.
\begin{theorem}
The \ac{AoA}, \ac{AoD}, and \ac{ToA} corresponding to path $p$ between \ac{RIS} subarray
$n$ and user $u$ are:\vspace{-.25cm}
\small
\begin{align}
\widetilde{\phi}^P_{n,u}=\arg \max _{\phi^P_{n,u}} \frac{\left|\widetilde{\boldsymbol{a}}_{k}^{H} \tilde{\boldsymbol{\zeta}}_{k}\left(\phi^P_{n,u}\right)\right|}{\left\|\widetilde{\boldsymbol{a}}_{k}\right\|_{2}\left\|\tilde{\boldsymbol{\zeta}}_{k}\left(\phi^P_{n,u}\right)\right\|_{2}}, \label{prop1}\\
\widetilde{\theta}^P_{n,u}=\arg \max _{\theta^P_{n,u}} \frac{\left|\widetilde{\boldsymbol{b}}_{k}^{H} \tilde{\boldsymbol{\xi}}_{k}\left(\theta^P_{n,u}\right)\right|}{\left\|\widetilde{\boldsymbol{b}}_{k}\right\|_{2}\left\|\tilde{\boldsymbol{\xi}}_{k}\left(\theta^P_{n,u}\right)\right\|_{2}},\label{prop2}\\
\widetilde{\tau}^P_{n,u}=\arg \min _{\tau^P_{n,u}} \frac{\left|\widetilde{\boldsymbol{c}}_{k}^{H} \tilde{\boldsymbol{\Lambda}}^P_{n,u,k}\left(\theta^P_{n,u}\right)\right|}{\left\|\widetilde{\boldsymbol{c}}_{k}\right\|_{2}\left\|\tilde{\boldsymbol{\Lambda}}^P_{n,u,k}\left(\tau^P_{n,u}\right)\right\|_{2}}.\label{prop3}
\end{align}
\normalsize
\vspace{-.25cm}
\begin{IEEEproof}
See Appendix A
\end{IEEEproof}
\end{theorem}
\eqref{prop1}-\eqref{prop3} can be solved with a one-dimensional search. Thus, the estimated \ac{AoA}, \ac{AoD}, and \ac{ToA} allow us to track the \ac{XR} user in near real-time. After obtaining $\widetilde{\phi}^P_{n,u}, \widetilde{\theta}^P_{n,u},$ and $\widetilde{\tau}^P_{n,u}$, the attenuation factor $\alpha^P_{n,u}$ can be obtained by substitution. To determine the tracking information of the user and the blockers, we use our approach in~Algorithm~1 in~\cite{chaccour2021joint}, which enables the \ac{RIS} characterize the spatial availability of communications. 
\subsection{Uniqueness and Accuracy of Estimated Sensing Parameters}
We now further analyze our tensor decomposition technique and our estimated sensing parameters. The accuracy and precision of our estimations play a fundamental role for the service needs of the \ac{XR} application and the \ac{THz} system performance. For example, in \ac{VR}, a precise tracking of the user's head and body movements is essential for creating a fully immersive experience. Meanwhile, in \ac{AR}, an accurate registration of virtual objects onto the real-world scene is crucial for providing an uninterrupted and continuous experience. Next, we prove that our obtained sensing parameters are unique. Then, we capture the accuracy of the sensing parameters via the \ac{CRLB} which is a fundamental theoretical limit on the precision of any unbiased estimator~\cite{kay1993fundamentals} that also provides a measure of the minimum acheivable variance in the estimation of a sensing parameter. Thus, knowing the \ac{CRLB} enables us to determine the accuracy and the sensitivity of our tensor decomposition mechanism in identifying the sensing parameters.
\subsubsection{Uniqueness of Sensing Parameters}
Achieving uniqueness in sensing parameters is essential in the Candecomp/Parafac (CP) decomposition-based method, which relies on Kruskal's condition~\cite{kruskal1977three}. Next, we prove the uniqueness of our solution.
\begin{lemma}
The \ac{XR} user's estimated tracking parameters, obtained at \ac{LoS}, $\{\widetilde{\phi}^L, \widetilde{\theta}^L, \widetilde{\tau}^L\}$, and the blocker's estimated sensing parameters, obtained at \ac{NLoS}, $\{\widetilde{\phi}^N, \widetilde{\theta}^N, \widetilde{\tau}^N \}$ are guaranteed to be unique because of the sparsity of the \ac{THz} channel.
\end{lemma}
\begin{IEEEproof}
	See Appendix B
\end{IEEEproof}
\subsubsection{\ac{CRLB} of Sensing Parameters}
We now derive the \ac{CRLB} as follows:
\begin{proposition}\label{Prop2}
The \ac{CRLB} of the \ac{XR} users sensing parameters at \ac{LoS} and \ac{NLoS} is given by:
\begin{equation}
\mathrm{CRLB}(\boldsymbol{S}^P)= \left[ \mathbb{E}\left\{\left(\frac{\partial L(\boldsymbol{S}^P)}{\partial \boldsymbol{S}^P}\right)^H\left(\frac{\partial L(\boldsymbol{S}^P)}{\partial \boldsymbol{p}}\right)\right\}\right] ^{-1},
\end{equation}
where $\boldsymbol{S}^P$ is the vector of sensing parameters, and $L(\boldsymbol{S}^P)=-JTK \log{(\pi\sigma_{\mathcal{W}}}) -\frac{1}{\sigma_{\mathcal{W}}} \norm{\boldsymbol{Y}^T_{(1)}-(C\odot B) A^T }_F$ is the log-likelihood of $\boldsymbol{S}^P$. Here, $Y_{(1)}$ is the mode-1 unfolding of $\mathcal{\chi}$ and  $\sigma_{\mathcal{W}}$ is the total noise standard deviation in the tensor domain.
\begin{IEEEproof}
The proof builds on~\cite{tichavsky2013, zhou2017low, kay1993fundamentals, xiangqian01}, and is omitted due to space constraints.
\end{IEEEproof}	
\end{proposition}
From Proposition~\ref{Prop2}, one can see that our estimation is mainly driven by the noise of the channel. Interestingly, given that we are operating at \ac{THz} bands, the total noise power is the sum of the molecular absorption noise and the Johnson-Nyquist noise generated by the thermal agitation of electrons in conductors. Thus, while sensing at higher carrier frequencies and larger bandwidths may achieve higher resolutions, \emph{the molecular absorption noise at higher carrier frequencies will affect the accuracy of the estimated sensing parameters}. With higher resolutions, it is necessary to control the variance of the estimates to considerably lower bounds\footnote{For instance, when the sensing parameters have a resolution in the centimeter range, the variance should be in the millimeter or micrometer range.}. Hence, there exists \emph{a duality between communication and sensing as we go to higher carrier frequencies}. Thus, the operation at \emph{sub-\ac{THz}} bands is increasingly appealing as it could provide high data rates and high-resolution sensing while circumventing the inherent unreliability of higher frequencies, which are further compromised by molecular absorption losses.
\section{Predicting Continual and Future User Behavior via Generative AI}\label{section:AI_Sensing}
Thus far, we have used tensor theory to analyze the received signal strength and extract the tracking parameters for \ac{LoS} links, and the mapping information for \ac{NLoS} links. Nonetheless, our sensing information from Section~\ref{section:tensor} is not comprehensive and is dependent on the user's alignment with the subarray. For instance, when all users are in perfect \ac{LoS} with their corresponding subarray, our Section~\ref{section:tensor} measurements cannot characterize the possibility of blockage that may or may not occur. Similarly, if a particular user is in \ac{NLoS}, our tensor-driven approach alone cannot enable tracking the user's DoF during these time slots. As a result, we would not be able to assess the \emph{comprehensive and continual user behavior and environment dynamics.} One could argue that such gaps in user behavior and environment dynamics can be found via \emph{averaging}, nonetheless, the stringent requirements of \ac{XR} services impose a degree of immersion which requires sensing information to be available \emph{instantaneously and continuously}. Here, we will first propose a novel non-autoregressive approach to predict the missing values of \ac{LoS} and \ac{NLoS} sensing matrices, namely \emph{our imputation system}\footnote{Imputation refers to the ability to intelligently recover and predict missing sequential sensing values to obtain a continuous and instantaneous sensing vector versus time.}. Then, we use these comprehensive time-variant vectors to predict future time slots via an encoder-decoder transformed based generative \ac{AI} framework. To the best of our knowledge, this is the first work that combines both imputation and forecasting systems in a single framework. This novel integration offers a significant advantage over existing methods by providing a more accurate and generalizable prediction of future time slots based on continuous and comprehensive sensing data. As will be evident later, such future user behavior predictions will enable us to control the uncertainty of the \ac{THz} channel, improve the overall performance instantaneously, and will enable the user  \emph{experience to be resilient}. Note that these predictions are not predicting the proper \ac{LoS} and \ac{NLoS} values for the time-slots with missing or future values. Instead, they provide \ac{RIS}-operated \acp{BS} instantaneous and continual information about the users' behavior and their corresponding varying environments. Thus, this \ac{E2E} imputation-forecasting learning framework enables us to obtain a comprehensive sensing cognizance in terms of tracking and situational awareness. 
\begin{figure}[t]
	\centering
	\includegraphics[scale=0.35]{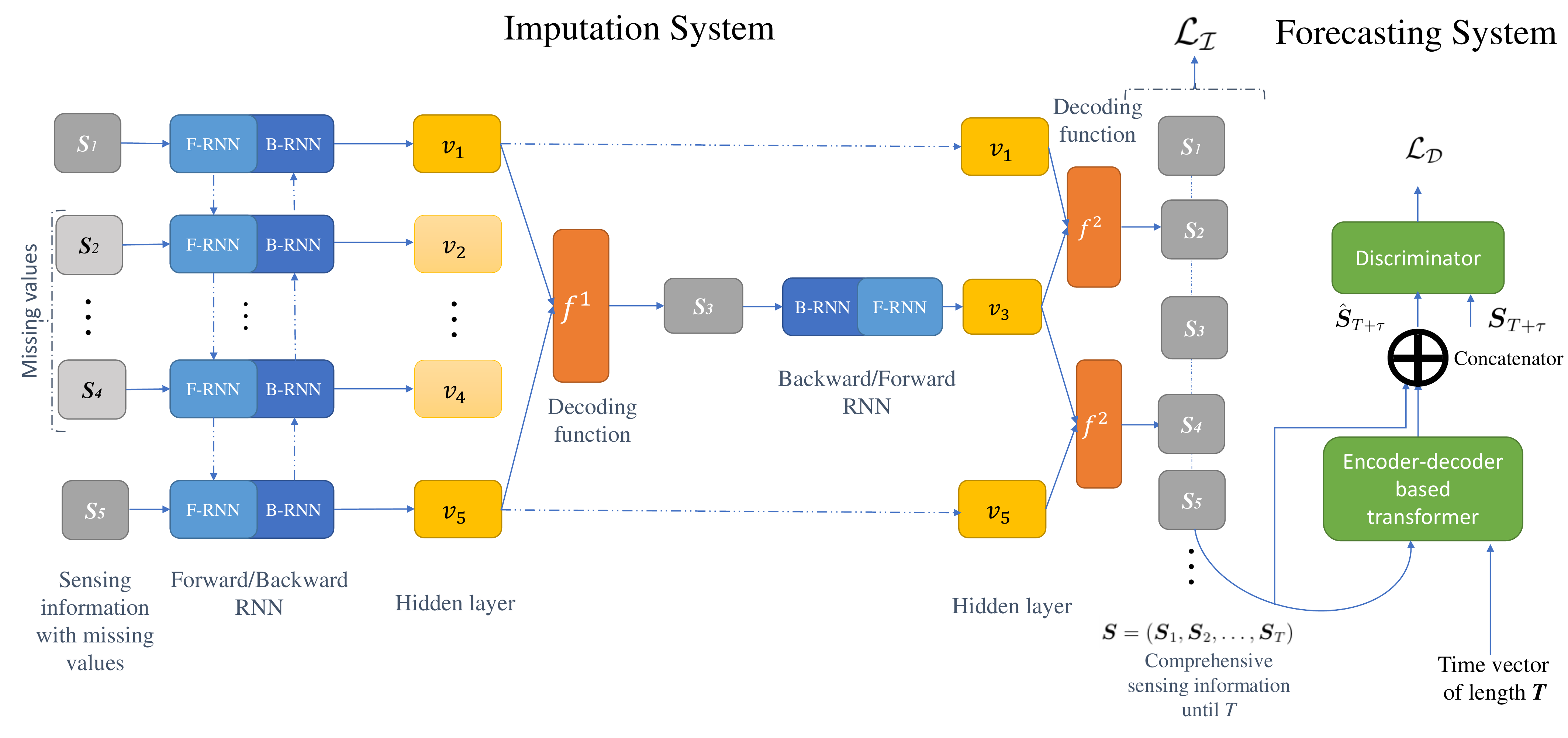}
	\caption{\small{Illustrative figure showcasing: A) The non-autoregressive multi-resolution imputation approach for predicting missing sensing information. For illustrative purposes the figure, we show a sequence of five sensing values, that include three missing values. B) The forecasting approach via an adversarial sparse transformer model. }}
	\label{fig:NAOMI}  
	\vspace{-.4cm}
\end{figure}
\subsection{Continual Prediction of User Behavior and Environmental Dynamics via Non-Autoregressive Multi-Resolution Generation}
Consider the link between user $u$ and \ac{RIS} subarray $n$. To simplify the notation, we omit subscripts $u$ and $n$. Let
$\boldsymbol{S}^L= (\boldsymbol{S}_{1}^L, \boldsymbol{S}_{2}^L, \dots, \boldsymbol{S}_{T}^L)$ be the sequence of $T$ snapshots of \ac{LoS} sensing information, whereby $\boldsymbol{S}_{i}^L= [\widetilde{\phi}^L, \widetilde{\theta}^L, \widetilde{\tau}^L ]^T$. We also define $\boldsymbol{S}^N= (\boldsymbol{S}_{1}^N, \boldsymbol{S}_{2}^N, \dots, \boldsymbol{S}_{T}^N)$ as the sequence of snapshots of \ac{NLoS} whereby $\boldsymbol{S}_{i}^N= [\widetilde{\phi}^N, \widetilde{\theta}^N, \widetilde{\tau}^N ]^T$. These values represent captured sensing information that characterizes the behavior of users and their corresponding dynamic time-varying enviornment. However, such vectors contain missing values due to beam misalignments and the intermittent nature of \ac{THz} links. Our goal is to characterize the \emph{continual and fine-grained} behavior of users and the time varying environment in these gaps, produced by beam misalignments, by predicting the missing values of $\boldsymbol{S}^{L}$ and $\boldsymbol{S}^{N}$. The missing values in $\boldsymbol{S}^{L}$ are captured by a masking sequence $\boldsymbol{\Xi}^L=(\boldsymbol{\Xi}_1^L, \boldsymbol{\Xi}_2^L, \dots, \boldsymbol{\Xi}_T^L)$,  while the missing values in $\boldsymbol{S}^{N}$ are found by its complement masking sequence $\Xi^N$. Thus, when the user is in \ac{LoS}, $\Xi^L_i=[1, 1, 1]^T$ and  $\Xi^N_i=[0, 0, 0]^T$. In contrast, in \ac{NLoS}, we have $\Xi^L_{i}=[0, 0, 0]^T$ and $\Xi^N_i=[1, 1, 1]^T$. We also define $p(\boldsymbol{S}^L_1, \dots, \boldsymbol{S}^L_T)= \prod_{t}{p(\boldsymbol{S}^L_t| \boldsymbol{S}^L_{<t})}$. To factorize this probability, one can use the chain rule and train a deep autoregressive model for imputation as in~\cite{liu2019naomi}. Nonetheless, autoregressive models perform this task using sub-optimal values when attempting to search the missing values and are susceptible to error compounding for sequences with a long range. Such autoregressive models assume that the current value of a time series is a function of its past values. As the length of the sequence increases, the number of required past values also increases, resulting in a very large number of parameters to estimate. Since we want to capture fine-grained sensing information, our measurements can consist of long sequences which can lead to overfitting and poor performance in capturing long-term dependencies. This creates a mismatch between the imputed sensing information that we must predict and the actual, observed indoor dynamics. To address these challenges, we propose a novel joint non-autoregressive generative and adversarial approach, whereby a deep non-autoregressive multi-resolution generative model~\cite{liu2019naomi} is used to recover missing observational sensing information. Meanwhile, an encoder-decoder based transformer~\cite{wu2020adversarial} uses the new continual predicted sensing information to predict future information.\\
\indent The architecture of this deep non-autoregressive, multi-resolution, generative model~\cite{liu2019naomi} consists of two components: a) a forward-backward encoder that maps the incomplete sequences of \ac{LoS} and \ac{NLoS} sensing sequences into hidden representations, and b) a multi-resolution decoder that imputes the missing values of \ac{LoS} given the hidden representations constructed in a). 
\subsubsection{Forward-backward encoder to represent missing sensing information}
Every $\boldsymbol{S}^L$, $\boldsymbol{S}^N$  and its corresponding mask $\boldsymbol{\Xi}^L$ and $\boldsymbol{\Xi}^N$  are concatenated as $\boldsymbol{\Theta}^L= [\boldsymbol{S}^L, \boldsymbol{\Xi}^L],$ and $\boldsymbol{\Theta^N}=[\boldsymbol{S}^N, \boldsymbol{\Xi}^N]$. The \ac{LoS} and \ac{NLoS} sensing information will undergo an encoder, which models a conditional distribution of two sets of hidden states given the sensing and masked input. We define $\Upsilon^{F,L}=(\upsilon_1^{F,L}, \dots, \upsilon_T^{F,L})$ and  $\Upsilon^{F,N}=(\upsilon_1^{F,N}, \dots, \upsilon_T^{F,N})$ as the forward hidden states as well as $\Upsilon^{G,L}=(\upsilon_1^{G,L}, \dots, \upsilon_T^{G,L})$ as the backward hidden states. Subsequently, the conditional probability distribution functions will be:
\begin{align}\label{conditionals}
p(\Upsilon^{F,P}|\Theta^{P})=\prod_{t=1}^{T} p(\upsilon_t^{F,P}| \upsilon_{<t}^{F,P}, \Theta^{P}_{\leq t}), \hspace{1cm}   p(\Upsilon^{G,P}|\Theta^{P})=\prod_{t=1}^{T} p(\upsilon_t^{G,P}| \upsilon_{>t}^{G,P}, \Theta^{P}_{\geq t}), 
\end{align}
where $P=\{L,N\}$ indicates the \ac{LoS} or \ac{NLoS} path. In~\eqref{conditionals}, we can see that the product of distributions can be parametrized via a forward and backward \ac{RNN}. 
\subsubsection{Multi-resolution decoder}
After representing the missing sensing information via forward and backward hidden state, namely $\Upsilon^P:=[\Upsilon^{G,P}, \Upsilon^{F,P}]$, the multi-resolution decoder must learn the distribution of the complete sensing sequence given such hidden states, i.e., $p(\boldsymbol{S}^P|\Upsilon^P) $. This a multi-resolution decoder can handle different time scales thereby improving the \emph{generalizability} of our prediction model with respect to the rate of change of the environment. That is, modeling different time scales, we can capture both short-term and long-term dependencies in the data, which can be important for accurate imputation. As in~\cite{liu2019naomi}, our multi-resolution decoder performs its decoding strategy from the most coarse to the finest-grained resolutions.
A decoder consisting of $Z$ resolutions is made up of decoding functions $f^1, \dots, f^Z$, each of which predicts every $\Delta_z=2^{Z-z}$ steps. To select the missing step $t$, which is close to the midpoint $[\frac{(i + j)}{2}]$, the decoder first finds two known steps $i$ and $j$ as pivots and, then, it determines the smallest resolution $z$ that satisfies $\Delta_z \leq \frac{(j-i)}{2}$. Using the forward states $\upsilon^F_i$ and the backward states $\upsilon^B_j$, the decoder updates the hidden states at time $t\star$. Subsequently, a decoding function $f^z$ is applied to map the hidden states to the distribution over the outputs, denoted as $p(S_i^{P \star}|\Upsilon^P)$. Once an imputation round is complete, the mask $\Xi^P$ is updated to $[1, 1, 1]^T$ and the process proceeds to the next resolution. 
\subsubsection{Imputation Approach Learning Objective}
Let $\mathcal{S}={\boldsymbol{S}\star}$ be the collection of the comprehensive sensing information (containing continual and uninterrupted sensing information). $\mathcal{I}_{\omega}(\boldsymbol{S}, \Xi)$ represents our generative imputation model parametrized by $\omega$, and $p(\Xi)$ is the prior over masking computed by averaging the blockage score over previous time slots. Thus, the system can be trained as follows:\vspace{-0.2cm}
\begin{equation*}\vspace{-0.2cm}
\min _\omega\mathbb{E}_{\boldsymbol{S}^* \sim \mathcal{S}, \Xi \sim p(\Xi), \hat{S} \sim \mathcal{I}_{(\boldsymbol{S}, \Xi)}}\left[\sum_{t=1}^T \mathcal{L}_\mathcal{I}\left(\hat{\boldsymbol{S}}_t, \boldsymbol{S}_t\right)\right],  \quad \text{where} \quad \mathcal{L}_I= \norm{\hat{\boldsymbol{S}}_t - \boldsymbol{S}_t}^2.\\
\end{equation*}
\indent Minimizing this objective function enables us to go from coarse grained sensing vectors with missing information in regards to user behavior towards fine-grained situational awareness which yields continual context in regards to the user behavior and the environmental time-varying dynamics until \emph{present time}. Next, we discuss our forecasting system.\vspace{-0.2cm}
 \subsection{Adversarial Transformer for Future User Behavior and Environment Forecasting}
 \subsubsection{Forecasting Model Architecture}
 So far, we obtained a comprehensive, complete, and continual sensing information (until present time) that empowers both the network and \ac{XR} application to instantaneously respond to user and environmental behaviors. To enable \emph{proactive anticipatory intelligence}, i.e., the ability to proactively anticipate and respond to user and environmental behaviors, we propose an encoder-decoder based transformer structure and an auxiliary discriminator, as in Fig.~\ref{fig:NAOMI}, as \emph{a forecasting system} for our continual sensing information. Encoder-decoder based transformers are known to be good for time series forecasting by virtue of their attention mechanisms in multi-head attention layers that allows the transformer to capture long-term dependencies within the continual sensing data outputed by our imputations. Similarly to the neural network layer structure of~\cite{wu2020adversarial}, the encoder and the decoder both contain an identical number of layers. Each layer consists of a multi-head self attention layer and a feed-forward network. For the transformer, given a novel architecture family of transformations, namely, $\alpha$-entmax (see~\cite{peters2019sparse, blondel2019learning}) was adopted\footnote{In essence, sparse attention models exhibit a form inductive bias that increases focus on relevant sensing information and makes the continual sensing sequence more interpretable as a result of the causality in the user's mobility.}.\\
\indent To improve the \emph{generalizability} of our \ac{AI} framework and properly characterize the stochasticity of the data, we adopt an adversarial training process. Rather than relying on minimizing likelihood or quantile loss functions, we use the adversarial training process to regularize the sensing data generated from the encoder-decoder network. By doing so, we are able to improve the robustness of our framework and better handle variations in the data. Adversarial learning here enables the framework to learn from both the observed and unobserved data, thus improving its ability to generalize to new, unseen data. Thus, per Fig.~\ref{fig:NAOMI}, a discriminator network is added to our forecasting system to improve the time-variant accuracy. Similarly to~\cite{wu2020adversarial}, the discriminator network consists of three fully connected linear layers with LeakReLu~\cite{xu2015empirical}. 
 \subsubsection{Forecasting Learning Objective}
 First, the goal of the discriminator is to distinguish whether the input sensing data are predicted or ground-truth values by computing the cross-entropy loss function. Here, the encoder-decoder transformer network acts as the generator $\mathcal{G}$ whose goal is to minimize the quantile loss and adversarial loss. Henceforth, the goal of the overall forecasting system is to optimize the following objective:
 \begin{equation}
 \begin{array}{c}
 \arg \min _{\mathcal{G}} \max _{\mathcal{D}} \lambda_\mathcal{A} \mathcal{L}_{a d v}\left(\omega^{\mathcal{G}}, \omega^{\mathcal{D}}\right)+\mathcal{L}_\rho\left(\omega^{\mathcal{G}}\right) \\
 \mathcal{L}_{a d v}\left(\omega^{\mathcal{G}}, \omega^{\mathcal{D}}\right)=\mathbb{E}\left[\log \left(\mathcal{D}\left(\boldsymbol{S}\right)\right]+\mathbb{E}\left[\log \left(1-\mathcal{D}\left(\hat{\boldsymbol{S}}\right)\right)\right]\right. \nonumber\\
 \mathcal{L}_\rho\left(\omega^{\mathcal{G}}\right)=2 \sum_{t=0}^\mathcal{T} \sum_{i=t_0+1}^{t_0+\tau} P_\rho\left(\boldsymbol{S}_{i, t}, \hat{\boldsymbol{S}}_{i, t}\right), \quad P_\rho\left(\boldsymbol{S}_{t, i}, \hat{\boldsymbol{S}}_{t, i}\right)=\Delta \boldsymbol{S}_{t, i}\left(\rho \mathbbm{1}_{\hat{\boldsymbol{S}}_{t, i}>\boldsymbol{S}_{i, t}}-(1-\rho) \mathbbm{1}_{\hat{\boldsymbol{S}}_{i, t} \leq \boldsymbol{S}_{i, t}}\right) \nonumber,
 \end{array}
 \end{equation}
 where $\omega^G$ and $\omega^D$ are the parameters of the generator and discriminator respectively, $\mathcal{T}$ is the overall time span, $P_\rho$ is the normalized quantile loss ($\rho$-risk), $\Delta \boldsymbol{S}_{i, t}=(\hat{\boldsymbol{S}}_{i,t}-\boldsymbol{S}_{i,t})$, and $\lambda_\mathcal{A}$ is the trade-off hyperparameter between $\mathcal{L}_{a d v}$ and $\mathcal{L}_\rho$. The network is able to learn how to generate more accurate and reliable predictions by minimizing the quantile loss and adversarial loss, as well as regularizing the sensing data through the discriminator. This improves the generalizability of the overall framework by allowing it to properly account for the stochasticity of the data. Essentially, this adversarial generative approach encourages the network to learn the underlying distribution of the data, rather than simply memorizing the training data. As a result, the network can better generalize to new, unseen data, which is crucial for accurate future prediction and subsequent robust optimization. \\
\indent Given the aforementioned imputation and forecasting systems, our logical next step is to feed this information to the communication system so as to minimize the handover costs and improve the \ac{THz} link resilience, as explained next.
\section{\ac{QoPE}-Centric and Sensing-Aware Handover Optimization}\label{optimization_section}
In this section, our goal is to leverage the comprehensive and predictive sensing information that we extracted so as to optimize the wireless \ac{XR} system performance \emph{instantaneously}. To do so, we must first define the \ac{QoPE} for \ac{XR} users over the reality-virtuality spectrum.
\subsection{\ac{QoPE} of \ac{XR} users}
In a \ac{THz} network, handover optimization, whether in indoor or outdoor settings, is paramount due to the dynamic and unpredicatable nature of the environment. Essentially, \ac{THz} links have a limited range, narrow beamwidth, and are sensitive to blockages (including self-blockage). Thus, while reseources such as bandwidth, resource blocks, or power are important, handover has an immediate impact on the user \ac{QoPE} as it directly affects the ability to maintain a LoS link~\cite{chaccour2020can}. Traditionally, handover schemes rely on a combination of parameters, which usually include \ac{SINR} thresholding, signal strength, distance, and \ac{QoS} indicators. While these parameters are effective in ensuring reliable connectivity in conventional services, they fail to capture the user's perspective or their \ac{QoPE}. Therefore, adopting existing handover schemes in personalized highly immersive experiences, such as XR applications, can disrupt the seamless user experience. Moreover, as narrow \ac{THz} beams are susceptible to micro-mobility and micro-orientation, such handover techniques are likely to result in frequent beam interruptions and a lack of continual connectivity, negatively impacting the overall network performance. Thus, we need more sophisticated handover schemes that consider user-centric metrics and the unique \ac{THz} characteristics. Consequently, to design an \ac{XR} user-centric handover scheme, we must first tractably define the \ac{QoPE} of \ac{XR} users over the reality-virtuality continuum from \ac{AR} all the way to \ac{VR}. 
\begin{definition}\vspace{-0.25cm}
	The \ac{QoPE} of an \ac{XR} user is given by:
	\begin{equation}
\beta= \left(\lambda_1 R_d +\lambda_2 R_w\right) P(\vartheta \leq  \lambda_3\vartheta_{t}). 
	\end{equation}
\end{definition}
Here, $\lambda_1, \lambda_2,$ and $\lambda_3$ $\in [0,1]$, while $R_d$ and $R_w$ are the downlink and uplink data rates respectively. $\vartheta$ is the \ac{E2E} delay that encompasses the transmission delay, the computing, queuing, and handover delays. To characterize this continuum, we define a parameter $\aleph$ as the reality-virtuality scaling factor. For fully virtual services (\ac{VR}), $\aleph=1$, meanwhile for services that overlay reality with augmented content (\ac{AR}), $\aleph=0$. Essentially, the \ac{QoPE} must depend on the level of immersion in the virtual world and thus on the defined scaling factor $\aleph$. Hence, $\lambda_1-\lambda_2 \propto c\aleph$ and $\lambda_3 \propto (1-c\aleph)$, where $c$ depends on the \ac{XR} generation and device requirements. For instance, for $\aleph=0$, $\lambda_1$= $\lambda_2$, as $\aleph$ increases, the gap between $\lambda_1$ and $\lambda_2$ increases. This relationship is a byproduct of the fact that, in \ac{AR} a bidirectional high data rate is needed to sustain a high \ac{QoPE}; meanwhile, in \ac{VR} the focus is only on the downlink data rate. Furthermore, as $\aleph$ increases from $0$ to $1$, $\lambda_3$ decreases. This is because \ac{VR} requires a higher level of immersiveness and haptic feedback, and thus require more stringent reliability measures. Meanwhile, \ac{AR} does not require full immersion. Hence, as shown in~\eqref{handover-equations}, we propose a handover triggering criteria that integrates the \ac{QoPE} of the user and its situational awareness. Essentially, if there exists a dynamic blocker that is getting closer to the associated user in the current/future time slot, a handover must occur.
\begin{equation}\label{handover-equations}
    \begin{cases}
\boldsymbol{S}^N_{b,n,u}(t) \xrightarrow{} \boldsymbol{S}^L_{b,n,u}(t) \land \mathrm{CRLB} \leq \mathrm{CRLB}^{\textrm{th}} & \text{initiate sensing-aware handover},\\
\boldsymbol{S}^N_{b,n,u}(t+1) \xrightarrow{} \boldsymbol{S}^L_{b,n,u}(t+1) \land \mathcal{L}_\mathcal{I} \leq \mathcal{L}_\mathcal{I}^{\textrm{th}} & \text{initiate sensing-aware proactive handover},\\
\beta_{u,n,b} (t)< \beta_{\textrm{th}} &\text{initiate \ac{QoPE}-centric handover},\\
\beta_{u,n,b}(t) \geq \beta_{\textrm{th}} & \text{maintain the current link}.
\end{cases}    
\end{equation}
 Essentially, as shown in \eqref{handover-equations}, our handover triggering scheme leverages our predicted continual and future user behavior as well as the environment dynamics so as to anticipate future beam misalignment. Furthermore, when a behavior is not anticipated our handover scheme instantaneously gets triggered when the \ac{QoPE} of the user drops. This enables a \emph{resilient}~\cite{robertresilience} decision making mechanism that can flexibly interact with the environment regardless of its dynamics and the accuracy of our estimated and predicted sensing parameters.\\
\indent Although our handover process is sophisticated, the \ac{XR} user experience can still be slightly degraded during this time, especially when a handover occurs due to \ac{QoPE} drops. To ensure optimal performance, our network optimization framework must discourage frequent handovers that increase \ac{E2E} delay and impact \ac{QoPE}. Therefore, we propose a user-centric handover cost and the total number of handovers per meter for a user $u$ as follows:
\begin{equation}\label{handover_cost}\vspace{-0.25cm}
\varrho_u= \min (h_u v_u \vartheta^h_u,1), \hspace{0.5cm} h_u=\frac{ \sum_{n \in \mathcal{N}} \sum_{b \in \mathcal{B}} \left\lbrace P \left[ x_{b,n,u}(t)x_{b,n,u}(t-1)=0\right]\right\rbrace}{D_u},
\end{equation}
where $x_{b,n,u}=1$, if user $u$ is associated to subarray $n$ of \ac{RIS} $b$, and $x_{b,n,u}=0$ otherwise, $v_u$ is the velocity of \ac{XR} user $u$, $D_u$ is the total distance traveled by user $u$, and $\vartheta^h_u$ is the handover time delay. 
\subsection{Handover Optimization and User Association Problem Formulation}
To guarantee a seamless experience to each and every \ac{XR} user simultaneously, our goal is to minimize the handover cost defined in~\eqref{handover_cost}, while guaranteeing a high \ac{QoPE} for every single user instantaneously. This optimization must also be cognizant of the estimated sensing parameters and their corresponding errors. In essence, performing handover based on erroneous sensing information will jeopardize the network performance. Hence, we first define the collective utility of all the active subarrays as follows:\vspace{-.1cm}
\begin{equation}
G(\boldsymbol{x})=  \lim_{\mathcal{T} \rightarrow \infty}\sum_t^\mathcal{T}\sum_{u\in \mathcal{U}}\sum_{n \in \mathcal{N}} \sum_{b \in \mathcal{B}} x_{b,n,u}(t)\beta_{u,n,b}(t)\times (1-\varrho(t)).
\end{equation}
\indent The goal of the \acp{RIS} is to find optimal handover strategies that maximize the collective team utility. Let $\pi_{b,n}(x_{b,u,n}(t)|x_{b,u,n}(t-1), \boldsymbol{S}_{b,n,u}^L(t, t+1), \boldsymbol{S}_{b,n,u}^N(t, t+1) )$ be the strategy of subarray $n$ in \ac{RIS} $b$ that is defined as the probability that the considered subarray serves user $u$ at time $t$, after successfully serving user $u$ at $t-1$. This probability is further conditioned on the available \ac{QoPE} $\beta_{u,n,b}(t)$  and the situational awareness of the current and future time slots $\boldsymbol{S}_{b,n}^L(t, t+1)$ and $\boldsymbol{S}_{b,n}^N(t, t+1)$. As such, the expected collective utility is given by:
\begin{equation*}
\bar{G}(\boldsymbol{\pi})=\sum_{\boldsymbol{x} \in \mathcal{X}}G(\boldsymbol{x}) \prod_{b \in \mathcal{B}}\prod_{n \in \mathcal{N}} \prod_{u \in \mathcal{U}}\pi_{b,n}(x_{b,u,n}(t)|x_{b,u,n}(t-1), \beta_{u,n,b}(t), \boldsymbol{S}^L(t, t+1), \boldsymbol{S}^N(t, t+1) ),
\end{equation*} where $\mathcal{X}$ is the set of all possible associations that subarray $n$ of \ac{RIS} $b$ can perform at time $t$.
We seek to find a handover policy that optimizes the data rate, continuity, and robustness over all the active \ac{THz} links of the \acp{RIS}'s subarrays. We thus, formulate the handover and user association problem of an \ac{RIS}-assisted \ac{THz} indoor network serving \ac{XR} users:
\begin{subequations}\label{problem}\vspace{-0.25cm}
\begin{align}
\max_{\boldsymbol{\pi}=\{\pi_1,\dots, \pi_{B\times N}\}} \quad & \bar{G}(\boldsymbol{\pi}),& \label{objective}\\
\textrm{s.t.} \quad & \lambda_1 R_{d,u}(t)+\lambda_2 R_{w,u}(t) \leq \lambda_1 R^{th}_{d}(t)+\lambda_2 R^{th}_{w}(t), \forall t \in (0, \infty), \forall u \in \mathcal{U}, \label{const1}\\
& P(\vartheta(t) \leq  \lambda_3\vartheta_{th}) \geq  \gamma_{th,\aleph} ,  \forall u \in \mathcal{U}, \label{const2}\\
\sum_{\boldsymbol{x} \in \mathcal{X}}\prod_{b \in \mathcal{B}}\prod_{n \in \mathcal{N}} &\prod_{u \in \mathcal{U}}\pi_{b,n}(x_{b,u,n}(t)|x_{b,u,n}(t-1), \beta_{u,n,b}(t), \boldsymbol{S}^L(t,t+1), \boldsymbol{S}^N(t, t+1) )=1, \label{const3}\\
0 &\leq \pi_{b,n}(x_{b,u,n}(t)|x_{b,u,n}(t-1), \beta_{u,n,b}(t), \boldsymbol{S}^L(t, t+1), \boldsymbol{S}^N(t, t+1)) \leq 1. & \label{const4}
\end{align}
\end{subequations} 
\eqref{objective} maximizes the time average of \ac{QoPE} of active users associated to subarrays, while taking into account the handover cost. \eqref{const1} and~\eqref{const2} take into account each user's downlink/uplink data rate and reliability explicitly. \eqref{const1} is the downlink/uplink condition that guarantees the satisfaction of the rate requirement based on the type of \ac{XR} service on the virtuality-reality spectrum. Similarly, \eqref{const2} is the  reliability condition that guarantees that the \ac{E2E} delay is less than a desired threshold. This \ac{E2E} delay considers the computing, queuing, transmission, and handover latency. \eqref{const3} and~\eqref{const4} are feasibility conditions. From~\eqref{problem}, one observes that the objective function is non-convex. Furthermore, \eqref{objective} exhibits multiple inter-dependent and correlated parameters such as the \ac{QoPE}, the \ac{E2E} delay, and the handover cost. Additionally, the binary association decision variable $x_{b,n,u}$ makes this problem difficult to solve via vanilla non-convex optimization frameworks. Optimally solving this problem, i.e., guaranteeing bidirectional rate, reliability, and low latency for all \ac{XR} simultaneously and instantaneously requires imposing restrictions or assumption on mobility, blockage, or the sensed environment. However, making such assumptions would not provide an accurate representation of the dynamic behavior of the \ac{THz} network and the way in which \ac{XR} users interact with it. Consequently, to address the highly varying, dynamic, and non-stationary nature of the \ac{THz} channel and \ac{XR} user behavior, we propose a novel semi-distributed multi-agent \ac{RL} framework that can fulfill the personal requirements of individual users while shrinking the gap between the best-case and worst-case performance. This framework enables agents to learn from their experiences, from the predicted and forecasted sensing information in Section~\ref{section:AI_Sensing}, and adapt their decision-making process based on the current/future network conditions and user requirements, resulting in a more efficient, resilient, and personalized performance.\vspace{-0.35cm}
\subsection{Semi-Distributed Multi-Agent RL for Robust Handover Optimization}
Given that the goal of~\eqref{problem} is to deliver a user-centric \ac{QoPE}, it cannot be solved through a fully centralized approach where a central server makes handover decisions without considering the specific needs of individual users. Due to the highly dynamic and non-stationary nature of the \ac{THz} channel and the diverse needs of individual users, achieving instantaneous reliability for all users at all times is challenging. We adopt a resilient and robust approach in our optimization solution to address the inherent uncertainties and variations in the system. Instead of solely focusing on maximizing the \ac{QoPE} or reliability, we aim to find mechanisms that enable a quick rebound from occasional dips in performance. By adopting such an approach, we ensure that the system remains adaptable and can function optimally under varying conditions and scenarios. Consequently, we propose a novel semi-distributed multi-agent \ac{RL} framework that brainstorms collaboratively yet executes individually. That is to address individual user's complex time varying requirements, every subarray will have its own neural network to prioritize decision-making for its tagged user. To account for the welfare of the overall network, a centralized feedback characterizing the \ac{QoPE} of all other users, will be communicated from the central \ac{MEC} server to the reward of each subarray and its tagged user. Thus, guaranteeing the maximization of the collective utility in~\eqref{objective}. \\
\indent In fact, the main goal when solving~\eqref{problem} is to find a policy for each subarray that can maximize the individual \ac{QoPE} requirements of its associated \ac{XR} user given the available sensing information, while maximizing the collective utility, i.e., focusing on individual \ac{QoPE} while not overshadowing the collective performance. Furthermore, the obtained continual and future sensing information may be erroneous and do not always reflect the ground truth of each subarray. That is, in~\eqref{handover-equations}, we can see that the handover process is a function of the accuracy of estimations and predictions, thus our environment is \emph{partially observable}. We next introduce the components of our semi-distributed multi-agent \ac{RL} framework:
\begin{itemize}
	\item \emph{Agents:} The agents are the subarrays in $\mathcal{N}$ of all the \ac{RIS}-operated \acp{RIS} in $\mathcal{B}$.
	\item \emph{Observations:} To ensure fair collaboration among subarrays, each agent observes its sensing information, the \ac{QoPE} of its user, and the sum of \ac{QoPE} of all active users\footnote{Such observations can be made available by measuring the uplink/downlink sum rates at the \ac{RIS}-operated \acp{BS}, and by performing a ping that enables measuring the \ac{E2E} delay and reliability. }. Thus, $\boldsymbol{o}_{b,n,u}(t)=[x_{b,n,u}(t-1), \boldsymbol{S}^L(t, t+1), \boldsymbol{S}^N(t, t+1), \beta_{u,n,b}(t), \sum_{b \in \mathcal{B} }\sum_{n \in \mathcal{N}}\sum_{i\neq u \in \mathcal{U}}x_{b,u,n}(t) \beta_{u,n,b}(t),  \rho_u ]$. The set of observations of all the subarrays of all the \acp{RIS} is $\mathcal{O}=\{\boldsymbol{O}_0,\boldsymbol{O}_1, \dots, \boldsymbol{O}_{t}\},$ where $\boldsymbol{O}_t=[o^t_{1,1}, \dots,o^t_{B,N}]$ are the states of the subarrays at time step $t$. This enables each subarray to make a decision based on its locally available measurements as well as the previous association performed.
	\item \emph{Action:} The action of each agent is to perform a handover decision or to maintain the currently active link. This decision requires modifying $x_{b,n,u}(t)$'s binary value and adjusting the beam of the considered subarray to the new or currently associated user, based on their tracking information in $\boldsymbol{S}^L(t), \boldsymbol{S}^N(t)$. These actions are given by $\boldsymbol{a}_t=[a_{1,1}(t), \dots, a_{B,N}(t)] \in \mathcal{A}$.
	\item \emph{Strategies:} The strategy of each subarray is the handover and user association strategy in~\eqref{problem}. 
	\item \emph{History:} The history of each agent is defined as the set of all measurements, observations, and actions collected up to time $t$ so as to enable each learning agent opportunistically learn the time-varying nature of the policy. $\mathcal{H}_{b,n}(t)=\{a_{b,n}(t), x_{b,n,u}(t-1), \boldsymbol{S}^L(t, t+1), \boldsymbol{S}^N(t, t+1), \beta_{u,n,b}(t), \sum_{i\neq u \in \mathcal{U}}\beta_{u,n,b}(t), \rho_u \}$. Here, it is important to note that while each subarray takes actions following its own user-centric strategy, from an \ac{RL} perspective, one might think they are unaware of the actions being made from other \acp{UE}. That said, $\boldsymbol{S}^L(t, t+1)$ provides a local situational awareness of the surrounding blockers as well as a future prediction of this local area. In other words, each agent is only cognizant of the actions of other agents that may affect them. The centralized feedback $\sum_{b \in \mathcal{B} }\sum_{n \in \mathcal{N}}\sum_{i\neq u \in \mathcal{U}}x_{b,u,n}(t) \beta_{u,n,b}(t)$ enables weighing each agent's contribution to the total welfare of the network.
	\item \emph{Reward:} The reward of our system is designed in a semi-distributed fashion. That is, the reward of each subarrary must measure the benefit of a selected action on: a) its current, established link, and b) the effect of the action on the overall network's robustness and resilience. Thus, we have: $r_{b,n}(t)= x_{b,n,u}(t)\beta_{u,n,b}(t)\times (1-\varrho_{b,n,u}(t)) + \zeta_{b,n} \sum_{i\neq u \in \mathcal{U}}\sum_{n \in \mathcal{N}} \sum_{b \in \mathcal{B}} x_{b,n,u}(t)\beta_{u,n,b}(t)$, where $\zeta_{b,n} \in (0,1)$ that can be tuned based on factors such as the a) the density of users in the network, b) the carrier frequency and the corresponding beamwidth, c) the level of cooperation needed to achieve desirable results. The accumulated discounted reward can be given by $R_{b,n}(t)=\sum_{t'=t}^{\mathcal{T} }\gamma^{t'-t} r_{b,n}(t)$ where $\gamma$ is the discount factor. This accumulated reward exemplifies maximizing the collective utility in~\eqref{objective} in an individualistic and instantaneous way that can still improve the collective performance. Here, $\gamma=0$ amplifies the current instantaneous performance only, while ignoring any lessons that can be leveraged from previous handover policy. Meanwhile, $\gamma=1$ would average out the return. Note that, the optimal objective of maximizing the user's instantaneous \ac{QoPE} requires incorporating past handover policies and user associations, which is why setting $\gamma=0$ would result in a suboptimal learning process.
 \end{itemize}
\indent To address the time-varying and non-stationarity nature of the \ac{THz} network as well as the complexity of the multi-agent partially observable process, every agent or subarray adopts a hysteretic deep recurrent Q-network~\cite{omidshafiei2017deep, sana2020multi}. In essence, each subarray $n$ as acts as quasi-independent learner (not fully independent as the centralized feedback depends on other agents) and maintains its own hysteretic deep recurrent Q-network. In contrast to overly-optimistic multi-agent \ac{RL} frameworks like vanilla distributed deep Q-learning, which tend to overlook low returns caused by teammates' exploratory actions and subsequently result in significant overestimation of Q-values in stochastic environments, hysteretic deep recurrent Q-networks can discern that low returns may stem from environmental stochasticity and should not be disregarded. This approach is particularly crucial for optimizing \ac{THz} networks, which are susceptible to extreme events like sudden blockages and molecular absorption, as well as micro-mobility and micro-orientation changes. By taking the dynamic and unpredictable nature of the environment into account, this approach allows for improved adaptation and optimization, ensuring that the reliability of the system is maintained and quickly restored following intermittent \ac{THz} links. This is particularly crucial in providing uninterrupted and dependable service for \ac{XR} users, whose needs may change over time, and guarantees a resilient user experience that can promptly recover from any disruptions. Moreover, the semi-distributed learning framework used in this approach ensures guaranteed convergence due to the adoption of asynchronous updates, where agents update their policy and value function independently and asynchronously. Although the optimality of the solution may not be guaranteed in highly dynamic and non-stationary environments like \ac{THz} networks, the quick adaptation of deep hysteretic networks enables the system to maintain robust and resilient performance in response to unpredictable user behavior.
Hysteretic deep recurrent Q-networks leverage two distinct learning rates to handle the complex dynamics of the learning process. The first learning rate, $\eta_1$, is used when the temporal difference (TD)-error is non-negative. In contrast, the second learning rate, $\eta_2$, is much smaller and is utilized to slow down the degradation of Q-values associated with previously positive experiences resulting from successful operations \cite{omidshafiei2017deep}. By implementing this approach, hysteresis is introduced, allowing subarrays to be more resilient against negative learning, exploration, and concurrent actions. This approach significantly improves the stability and robustness of the learning process, leading to better overall performance and expedited convergence.\\
\indent Similarly to~\cite{sana2020multi}, the deep hysteretic Q-network adopted is with one input layer, two fully connected hidden layers, one \ac{RNN} hidden layer, a dueling layer, and an output layer. The subarray's local observations and the estimated state-action value $Q_{b.n}$, respectively define the input and output layer of the hysteretic deep recurrent Q-network. Multi-agent \ac{RL} frameworks are known to face the challenge of \emph{shadowed equilibria}, a phenomenon where local observations and non-stationarity cause locally optimal actions to become suboptimal at the global level~\cite{fulda}. Addressing this challenge effectively and stabilizing the learning process is crucial. To ensure a resilient policy for each subarray and enable a smooth recovery from sudden disruptions of THz links, we stabilize the learning process using a synchronized sampling strategy known as concurrent experience replay trajectories~\cite{omidshafiei2017deep}.
\vspace{-0.25cm}
\section{Simulation Results and Analysis}
For our simulations, we consider an indoor area modeled as a square of size $\SI{24}{m}\times\SI{24}{m}$ whereby the \acp{RIS} are deployed over its three walls. The molecular absorption was obtained from the sub-\ac{THz} model in \cite{kokkoniemi2021line} with $1\%$ of water vapor molecules. We set: $N=\SI{64}{}$ antennas, $Q=\SI{32}{}$ antennas, $f=\SI{0.275}{THz}$, $W=\SI{10}{GHz}$ (unless stated otherwise), and  $p=\SI{30}{dBm}$. All statistical results are averaged over a large number of independent runs. For \ac{AR}, we set $\lambda_1=1, \lambda_2=1,$ and $\lambda_3=1$. For \ac{VR}, we set $\lambda_1=1, \lambda_2=0.6,$ and $\lambda_3=0.8$, and, finally for \ac{MR}, we let  $\lambda_1=0.9, \lambda_2=0.9,$ and $\lambda_3=0.95.$ The reliability threshold set $\nu_t=\SI{24}{ms}$. \ac{VR} services ensure the immersion in a virtual world by guaranteeing the a \ac{QoPE} below the $\SI{20}{ms}$ as $\lambda_3=0.8$ for \ac{VR}. The network was simulated with data generated from \ac{XR} users moving according to a random walk scheme as well as a WiFi positioning dataset~\cite{ashraf2022wi} so as to have a general scheme for user mobility.\\
\begin{figure}
	\begin{minipage}{0.49\textwidth}
		\vspace{-.1cm}
		\centering
		\includegraphics[scale=0.150]{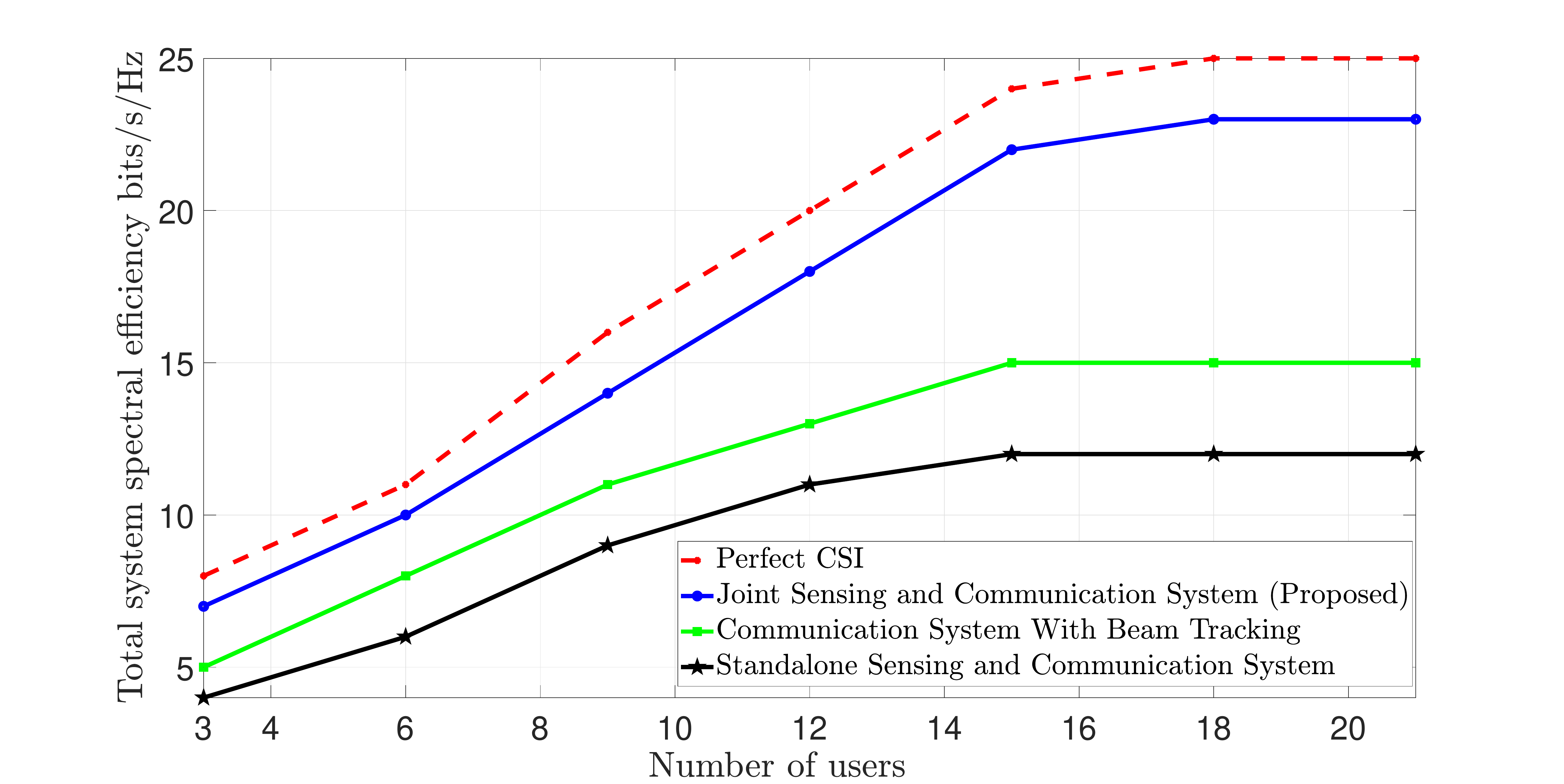}
		\vspace{-.20cm}
		\caption{\small{Spectral efficiency versus number of users.}}
		\label{fig:SpectEff} 
		\vspace{-0.7cm} 
	\end{minipage}
	\begin{minipage}{0.49\textwidth}
		\vspace{-.1cm}
		\centering
		\includegraphics[scale=0.3]{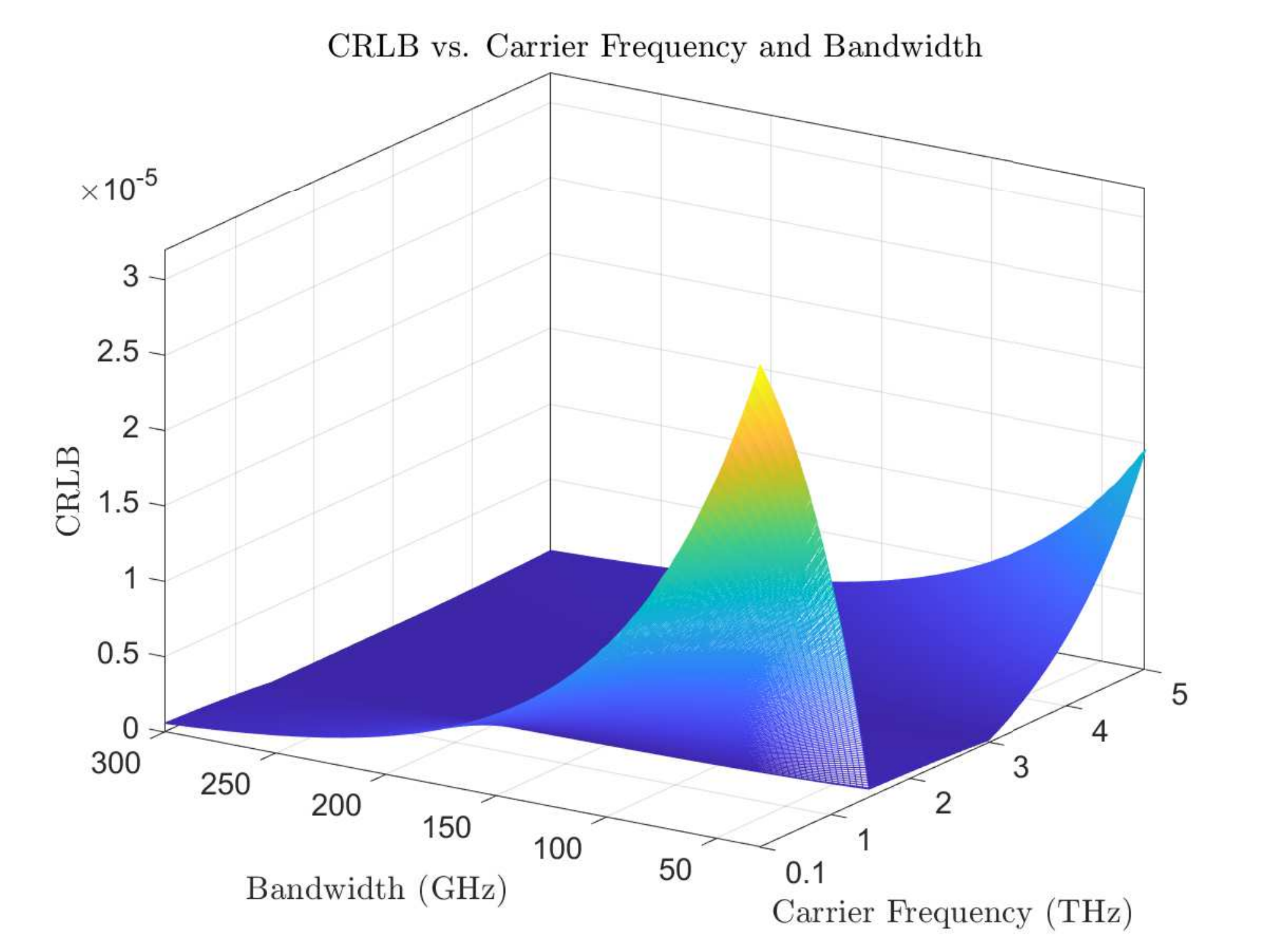}
		\vspace{-.20cm}
		\caption{\small{CRLB versus carrier frequency (THz) and bandwidth (GHz).}}
		\label{fig:CRLB} 
		\vspace{-0.7cm} 
	\end{minipage}
\end{figure}
\begin{figure}[t]
	\begin{minipage}{0.49\textwidth}
		\centering
		\includegraphics[scale=.30]{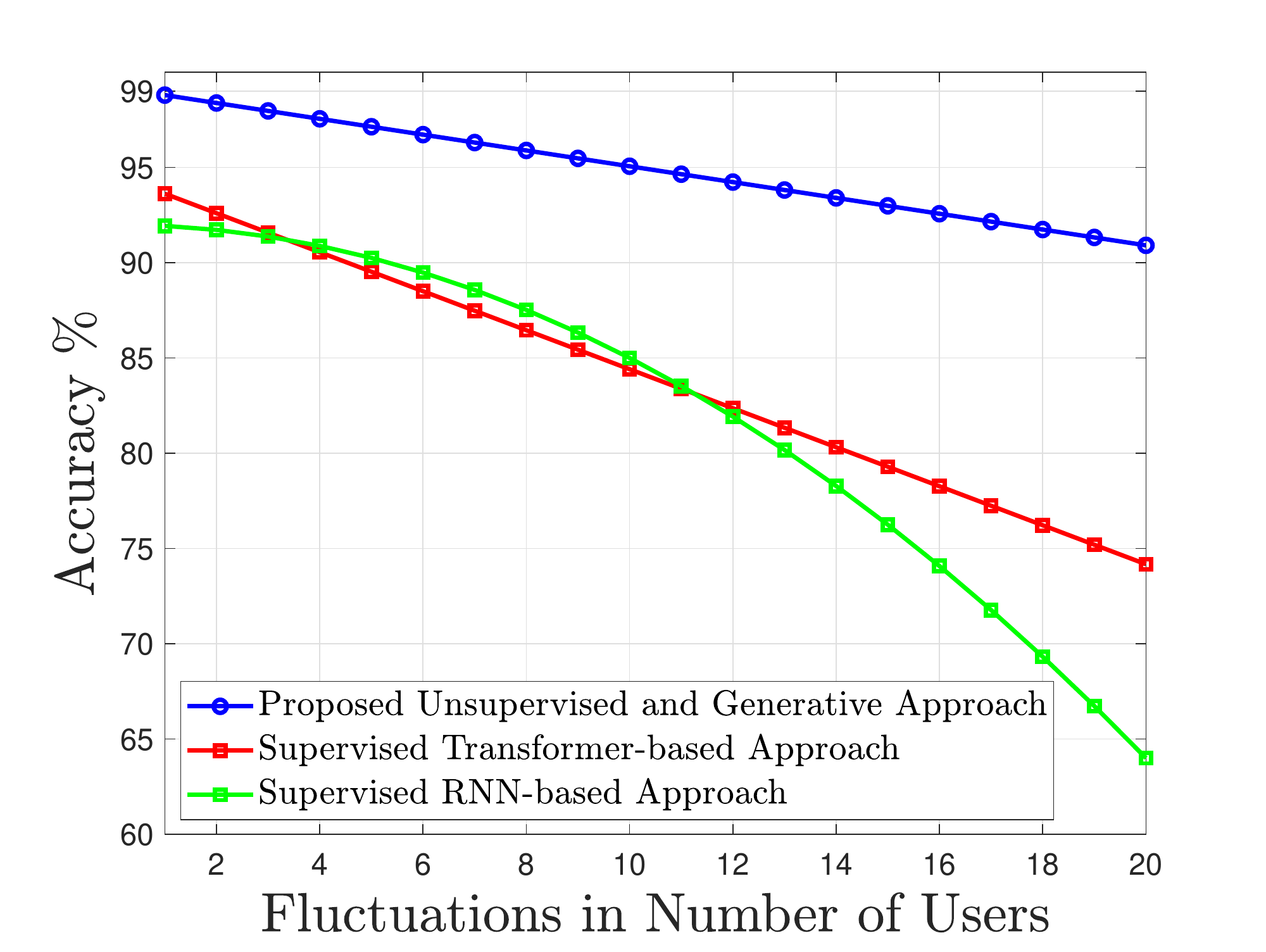}
		\vspace{-.20cm}
		\subcaption{}
		\label{fig:gen_users} 
		\vspace{-0.7cm} 
	\end{minipage}
	\begin{minipage}{0.49\textwidth}
		\centering
		\includegraphics[scale=0.30]{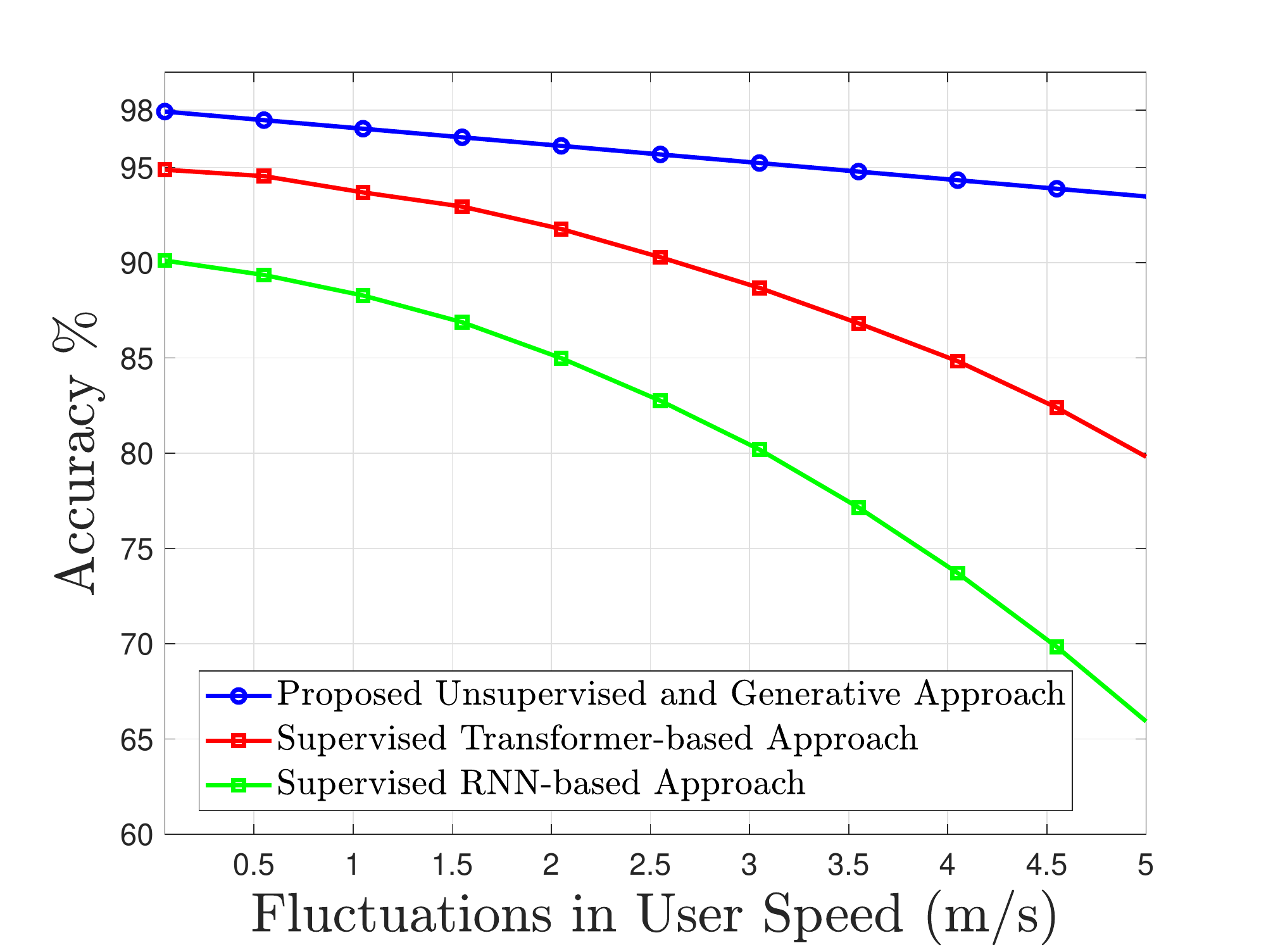}
		\vspace{-.20cm}
		\subcaption{}
		\label{fig:gen_velocity} 
		\vspace{-0.7cm} 
	\end{minipage}	
	\caption{\small{ Generalizability of our integrated imputation and forcesating system with respect to (a) Fluctuations in the number of users in/out of the indoor area, and (b) Fluctuations in user's speed (m/s).}} \label{figs:generalizability}
			\vspace{-.5cm} 
\end{figure}
\begin{figure}[t]
	\begin{minipage}{0.49\textwidth}
		\centering
		\includegraphics[scale=0.30]{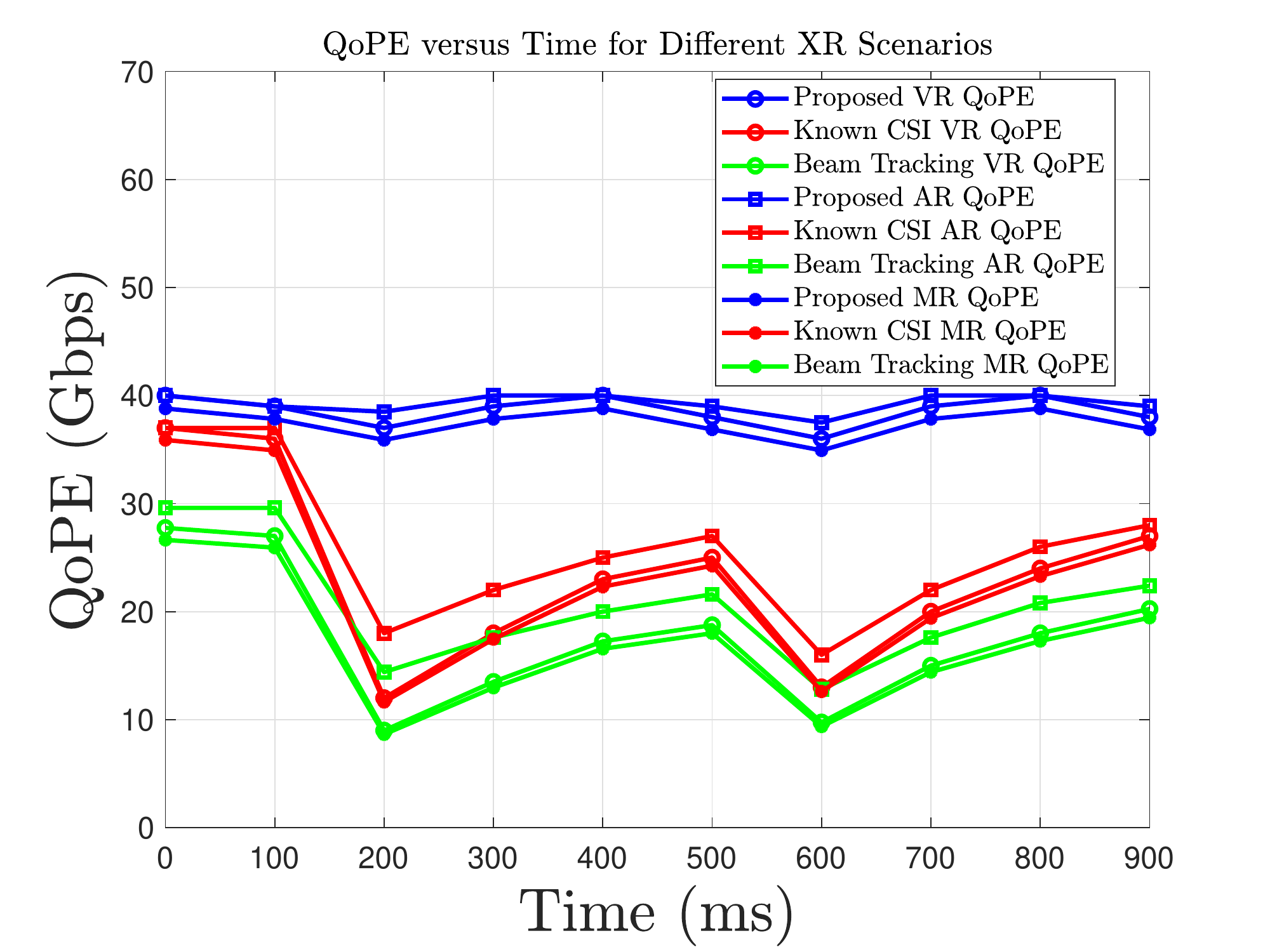}
		\subcaption{}
		\label{fig:QoPE_Session} 
		\vspace{-0.7cm} 
	\end{minipage}	
	\begin{minipage}{0.49\textwidth}
		\centering
		\includegraphics[scale=0.30]{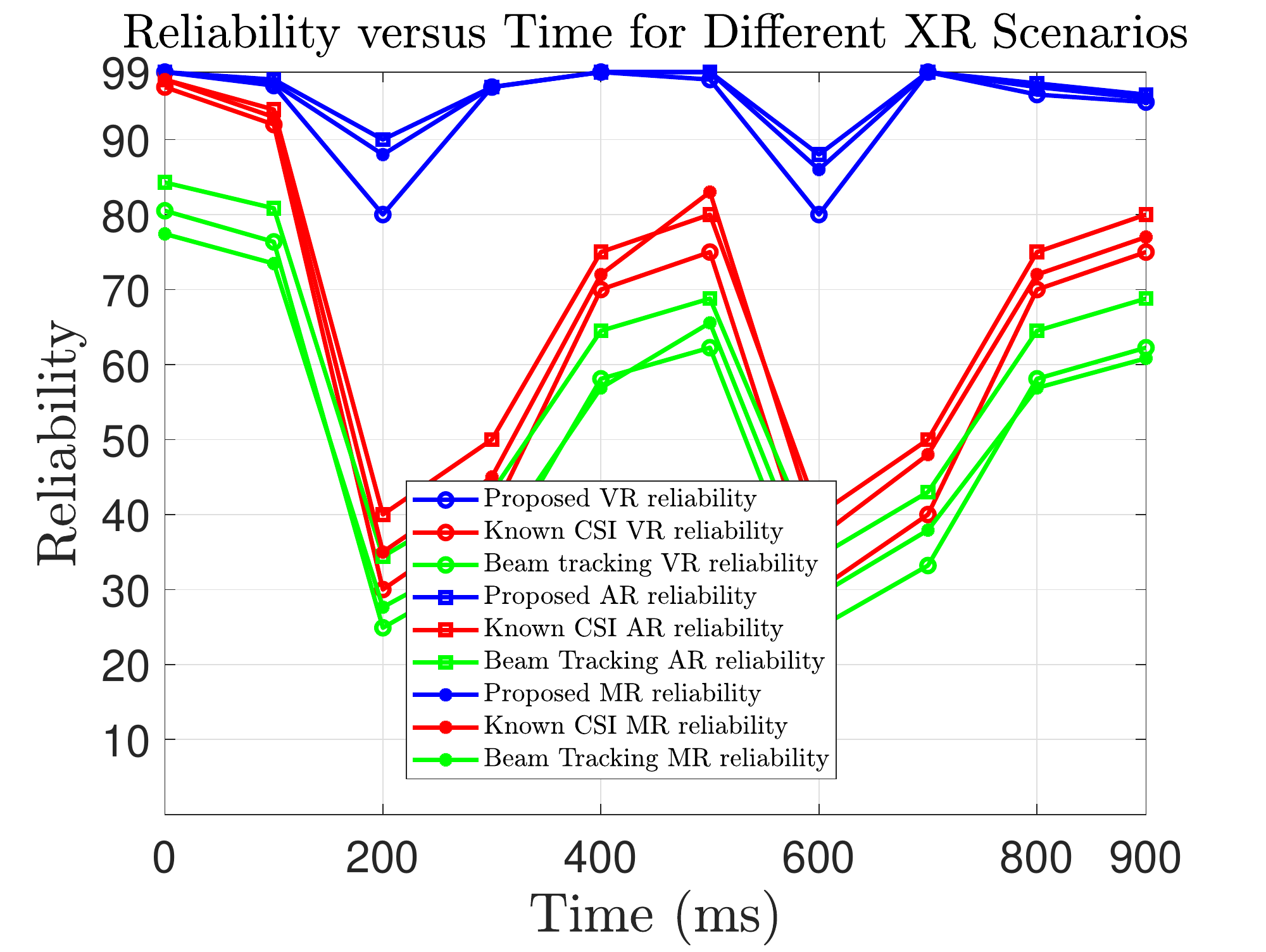}
		\vspace{-.20cm}
		\subcaption{}
		\label{fig:Rel_Session} 
		\vspace{-0.5cm} 
	\end{minipage}
	\caption{\small{(a) QoPE of XR users versus XR session time, (b) Reliability of XR users versus XR session time.}}		\vspace{-0.75cm} 
\end{figure}
\begin{figure}[t]
	\vspace{-.1cm}
	\centering
	\includegraphics[scale=0.35]{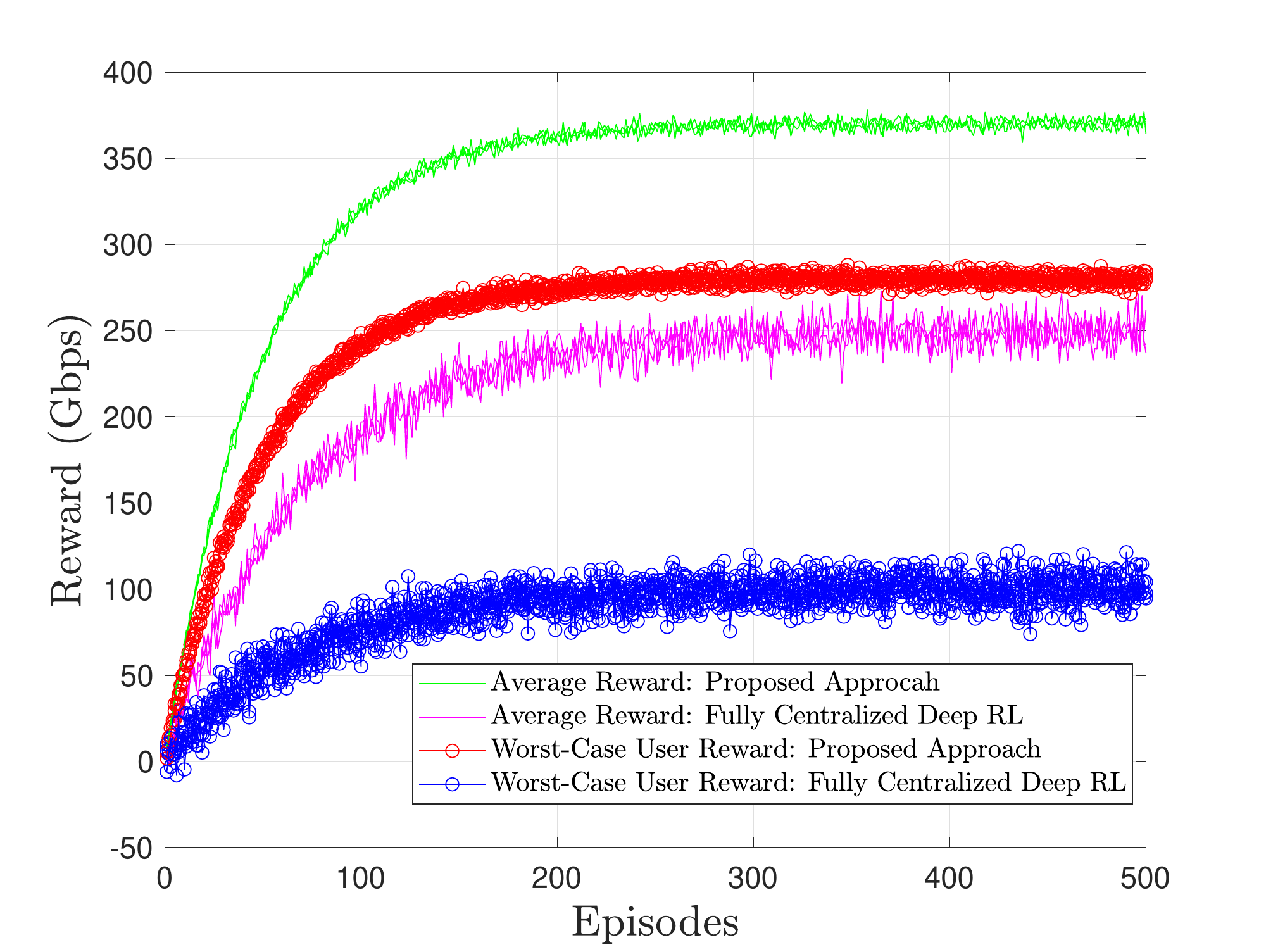}
	\vspace{-.1cm}
	\caption{\small{Reward versus online training episodes }}
	\label{fig:Reward} 
	\vspace{-.5cm} 
\end{figure}
\indent In Fig.~\ref{fig:SpectEff}, the total spectral efficiency defined as in \cite{chiriyath2017radar} is shown. Fig.~\ref{fig:SpectEff} clearly shows the benefits of deploying a joint sensing and communication system that shares hardware, waveform, and spectrum. In fact, our proposed joint sensing and communication approach achieves a $\SI{42}{\%}$ and $\SI{75}{\%}$ improvement respectively compared to a communication system with beam training and a standalone sensing and communication system. Fig.~\ref{fig:CRLB} showcases the effect of varying the bandwidth and the carrier frequency on the CRLB. We can see that as intuitively expected, higher bandwidths minimize the CRLB, nonetheless, the CRLB's behavior versus carrier frequencies is not monotonic. In fact, we can see that the CRLB decreases for carrier frequencies in the range of $\SI{0.1}{THz}-\SI{1}{THz}$. Nonetheless, for higher carrier frequency ranges, Fig.~\ref{fig:CRLB} shows that the CRLB starts to increase again due to the effect of more significant molecular absorption lines. From Fig.~\ref{fig:CRLB}, we also observe that increasing the bandwidth at such high carrier frequencies ($\SI{1}{THz}-\SI{5}{THz}$) can also enable decreasing the CRLB.\\
\indent In Fig.~\ref{figs:generalizability}, the generalizability of our joint imputation and forecasting system is evaluated versus fluctuations and new unseen user behavior (which are new mobility patterns tuned in this testing period). We compare our approach to \ac{RNN}-based and transformer-based approaches which are well known models for predicting temporal dynamics~\cite{li2023towards}. From Fig.~\ref{figs:generalizability}, we observe that, when increasing the fluctuations in the number of users as well as the users speed, the accuracy of our proposed scheme remains high. Meanwhile, deep learning learning schemes relying on datasets and having a bias cannot adapt to fluctuations properly. Their accuracy decreases rapidly as such fluctuations and unseen behavior pattern increases.\\ 
\indent In Fig.~\ref{fig:QoPE_Session}, the \ac{QoPE} of an \ac{XR} user (over different \ac{XR} scenarios on the continuum) is evaluated. We compare our proposed approach with communication-only schemes that use known \ac{CSI} and beam tracking so as to study the potential benefits of joint sensing and AI-driven situational awareness in enhancing communication system. First, we observe that maintaining a high \ac{AR} \ac{QoPE} is much easier than \ac{MR} and \ac{VR}. The difference in \ac{QoPE} increases because of extreme events such as blockages around time $\SI{200}{ms}$ and $\SI{600}{ms} $. Fig.~\ref{fig:QoPE_Session} shows that our proposed approach is able to recover easily from blockages and thereby demonstrating a \emph{high resilience}. This is by virtue of the network optimization design as well as the knowledge of future time slots. Meanwhile, for communication only schemes with known \ac{CSI} and beam tracking, the time needed to recover from such extreme events is much longer (by $\SI{270}{ms}$). In fact, by virtue of our semi-distributed, sensing-aware, and \ac{QoPE}-centric approach, our proposed approach achieves gains of $\SI{60.27}{\%}$ and $\SI{110}{\%}$ when compared to known \ac{CSI} and beam tracking schemes respectively. We observe a similar behavior for the reliability in Fig.~\ref{fig:Rel_Session}.  Indeed, the reliability of \ac{VR} (compared to \ac{AR} and \ac{MR}) is more difficult to maintain, as it requires a more stringent \ac{E2E} delay bound. Furthermore, similar to the \ac{QoPE} we can see that the reliability of the system takes time to recover from disruption in links in the known CSI and beam tracking schemes. Fig.~\ref{fig:Rel_Session} shows that, our proposed approach achieves $\SI{61}{\%}$ and $\SI{78}{\%}$ gains in the reliability compared to known \ac{CSI} and beam tracking schemes. \\
\indent In Fig.~\ref{fig:Reward}, the average reward and the worst-case user reward of our proposed approach are evaluated and compared to fully centralized deep \ac{RL} schemes. From Fig.~\ref{fig:Reward}, we observe that, for the average and the worst-case user, our semi-distributed deep Q hysteretic \ac{RL} scheme achieves higher rewards than fully a centralized deep \ac{RL} schemes. In fact, we can see that the worst-case reward of our proposed approach is higher than the average-reward of centralized deep \ac{RL} schemes. When comparing with centralized vanilla deep \ac{RL} frameworks, we observe a 1.5-fold improvement in the average reward and a two-fold improvement in the worst-case reward.
\vspace{-0.5cm}
\section{Conclusion}
In this paper, we have proposed a novel joint sensing, communication, and AI approach for achieving optimized and resilient wireless XR experiences at THz bands. We leverage THz-operated RISs as base stations to extract sensing parameters by opportunistically utilizing uplink communication waveforms. Our framework comprises a tensor decomposition framework, a non-autoregressive multi-resolution generative AI framework with an adversarial transformer, and a QoPE-centric and sensing-driven optimization that uses a semi-distributed multi-agent deep recurrent hysteretic Q-neural network. Our goal is to maximize individual QoPEs and improve the robustness and resilience of THz links. Through our analysis and simulations, we have concluded the following observations:
\begin{itemize}
	\item Deploying THz base stations at carrier frequencies higher than $\SI{1}{THz}$ is challenging in the sensing and communication realm. Although such bands may offer high bandwidths for high-resolution sensing and data rates, our results show that the possibility of performing erroneous sensing estimates increases due to molecular absorption lines. Hence, there is a need for open problems that enable higher rates at the sub-THz spectrum~\cite{chaccour2021seven}. 
	\item Ensuring an instantaneous reliability of five nines with \ac{THz} frequencies alone remains challenging even with sensing. However, by combining sensing and \ac{AI}, we can better control future networks and make them more robust and resilient. Further research is needed to explore the impact of joint sensing, communication, and AI in a multi-band scheme that integrates sub-$\SI{6}{GHz}$, \ac{mmWave}, and \ac{THz} frequencies to achieve resilient five nines reliabilities and high quality of personalized experience.
	\item Novel intelligence-centric metrics similar to the \ac{QoPE} should be proposed to ensure the requirements of future 6G and beyond applications are met. The resilience of certain application metrics must be evaluated with respect to time, specifically how quickly they rebound to their acceptable range. This is especially important for future 6G applications that require instantaneous performance.
\end{itemize}
\appendix
\vspace{-0.2cm}
\subsection{Proof of Theorem 1}
Problem \eqref{optimization} can be solved efficiently by the alternating least squares matrix factorization procedure \cite{sidiropoulos2017tensor}. Since we deal with a third order tensor, this factorization has three iterative steps. Then, it fixes two factor matrices and minimizes the error with respect to the considered factor matrix as:
\begin{equation*}
\small
\widetilde{\boldsymbol{A}}^{(t+1)}=\arg \min _{\widetilde{\boldsymbol{A}}}\left\|\boldsymbol{Y}_{n,(1)}^{T}-\left(\widetilde{\boldsymbol{C}}^{(t)} \odot \widetilde{\boldsymbol{B}}^{(t)}\right) \widetilde{\boldsymbol{A}}^{T}\right\|_{F}^{2}, \quad \small
\widetilde{\boldsymbol{B}}^{(t+1)}=\arg \min _{\widetilde{\boldsymbol{B}}}\left\|\boldsymbol{Y}_{n,(2)}^{T}-\left(\widetilde{\boldsymbol{C}}^{(t)} \odot \widetilde{\boldsymbol{A}}^{(t+1)}\right) \widetilde{\boldsymbol{B}}^{T}\right\|_{F}^{2}, 
\end{equation*}
\begin{equation}
\small
\widetilde{\boldsymbol{C}}^{(t+1)}=\arg \min _{\widetilde{\boldsymbol{C}}}\left\|\boldsymbol{Y}_{n,(3)}^{T}-\left(\widetilde{\boldsymbol{B}}^{(t+1)} \odot \widetilde{\boldsymbol{A}}^{(t+1)}\right) \widetilde{\boldsymbol{C}}^{T}\right\|_{F}^{2},
\end{equation}
where, $\odot$ is the Khatri-Rao product symbol. The exact factor matrices are related to their estimates according to:
$\widetilde{\boldsymbol{A}} =\boldsymbol{A} \boldsymbol{J}_{1} \boldsymbol{\Gamma}+\boldsymbol{E}_{1}, \hspace{0.1cm}
\widetilde{\boldsymbol{B}} =\boldsymbol{B} \boldsymbol{J}_{2} \boldsymbol{\Gamma}+\boldsymbol{E}_{2},  \hspace{0.1cm}
\widetilde{\boldsymbol{C}} =\boldsymbol{C} \boldsymbol{J}_{3} \boldsymbol{\Gamma}+\boldsymbol{E}_{3}, $
where $\{\boldsymbol{J}_1, \boldsymbol{J}_2, \boldsymbol{J}_3\}$ are unknown nonsingular diagonal matrices that satisfy $\boldsymbol{J}_1\boldsymbol{J}_2\boldsymbol{J}_3=
\boldsymbol{I}$, $\boldsymbol{\Gamma}$ is an unknown permutation matrix, and $\{\boldsymbol{E_1},\boldsymbol{E_2},\boldsymbol{E_3}\}$  are estimation error matrices. By applying a maximum likelihood estimator on each one of the equations and assuming that the estimation error matrices $\{\boldsymbol{E_1},\boldsymbol{E_2},\boldsymbol{E_3}\}$ follow an \ac{i.i.d.} circularly symmetric Gaussian distribution~\cite{zhou2017low}, we obtain \eqref{prop1}, \eqref{prop2}, and \eqref{prop3}. \vspace{-0.35cm}
\subsection{Proof of Lemma 1}
Proving the uniqueness of our sensing parameters requires proving that the CP decomposition of $\boldsymbol{\chi}$ is unique under mild conditions with scaling and permutation ambiguities. In essence, the Kruskal's condition~\cite{bhaskara2014uniqueness,kruskal1977three} guarantee the CP decomposition if the following condition is met: $k_A+ k_B+k_C \geq 2P +2 $,
where $k_A$, $k_B$, and $k_C$ are the Kruskal ranks of matrices $\boldsymbol{A}, \boldsymbol{B}, \boldsymbol{C}$ and $P$ is the number of \ac{THz} links. In fact, the Kruskal-rank of a matrix is defined the maximum number of any linearly independent columns that can be identified in a matrix. Based on~\cite{kruskal1977three}, $k_A$ and $k_B$ also denote the upper bound on the maximum number of links that can be distinguished. Hence, when this upper bound condition is satisfied, one can write the Kruskal-rank of the matrices $\boldsymbol{A}$, $\boldsymbol{B}$, $\boldsymbol{C}$: $
k_A= \min(J, P), \hspace{0.5cm} k_B=\min(T,P), \hspace{0.5cm} k_C=\min(K,P)$.
As we are operating in the \ac{THz} frequencies, the number of links available is quite limited, primarily restricted to either \ac{LoS} or \ac{NLoS}. However, the number of antennas and subcarriers at our disposal is significantly large. Consequently, the Kruskal condition is fulfilled, which ensures the uniqueness of our solution.
\bibliographystyle{IEEEtran}
\def\baselinestretch{0.85}
\bibliography{bibliography}
\end{document}

%% file: acronyms.tex
\acro{D2D}{device-to-device}
\acro{SIR}{signal-to-interference-ratio}
\acro{SINR}{signal-to-interference-plus-noise-ratio}
\acro{PCP}{Poisson cluster process}
\acro{CoMP}{coordinated multi-point}
\acro{BS}{base station} 
\acro{MD-CoMP}{macrodiversity CoMP transmission}
\acro{MAC}{medium-access-control}
\acro{JT-CoMP}{joint transmission CoMP}
\acro{CoMP-JT}{coordinated multipoint joint transmission}
\acro{SBS}{small base station}
\acro{MDSD}{multiple devices to a single device}
\acro{MDS}{maximum distance separable}
\acro{SCN}{small cell network}
\acro{PPP}{Poisson point process}
\acro{TCP}{Thomas cluster process}
\acro{CSI}{channel state information}
\acro{PDF}{probability distribution function}
\acro{PMF}{probability mass function}
\acro{RV}{random variable}
\acro{i.i.d.}{independently and identically distributed}
\acro{MBMS}{multimedia broadcasting multicasting service}
\acro{EE}{energy efficiency}
\acro{HCP}{hard-core placement}
\acro{CCDF}{complementary cumulative distribution function}
\acro{CDF}{cumulative distribution function}
\acro{PC}{probabilistic caching}
\acro{RC}{random caching}
\acro{CPF}{caching popular files} 
\acro{PGFL}{probability generating functional}
\acro{KKT}{Karush-Kuhn-Tucker}
\acro{PGF}{point generating function}
\acro{SCA}{successive convex approximation}
\acro{HD}{high-definition}
\acro{FHD}{full-high-definition}
\acro{UHD}{ultra-high-definition}
\acro{VR}{virtual reality}
\acro{AR}{augmented reality}
\acro{5G}{fifth-generation}
\acro{QoS}{quality-of-service}
\acro{QoE}{quality-of-experience}
\acro{IoT}{internet of things}
\acro{MHCPP}{Matern hardcore point process}
\acro{LoS}{line-of-sight}
\acro{NLoS}{non-line-of-sight}
\acro{PSD}{power spectral density}
\acro{MEC}{mobile edge computing}
\acro{E2E}{end-to-end}
\acro{THz}{terahertz}
\acro{CLT}{central limit theorem}
\acro{HQ}{High Quality}
\acro{eMBB}{enhanced mobile broadband}
\acro{URLLC}{ultra reliable low latency communications}
\acro{mmWave}{millimeter wave}
\acro{EVT}{extreme value theory}
\acro{GEV}{generalized extreme value}
\acro{LIS}{large intelligent surface}
\acro{RIS}{reconfigurable intelligent surface}
\acro{RF}{radio frequency}
\acro{UE}{user equipment}
\acro{MIMO}{multiple-input multiple-output}
\acro{EVaR}{entropic value-at-risk}
\acro{DNN}{deep neural network}
\acro{MDP}{Markov decision process}
\acro{RL}{reinforcement learning}
\acro{RNN}{recurrent neural network}
\acro{ANN}{artificial neural networks}
\acro{LSTM}{long short-term memory}
\acro{ReLu}{rectified linear unit}
\acro{VaR}{value-at-risk}
\acro{SNR}{signal-to-noise ratio}
\acro{AoSA}{array of subarray}
\acro{XR}{extended reality}
\acro{AoA}{angle of arrival}
\acro{ULA}{uniform linear array}
\acro{AoD}{angle of departure}
\acro{EM}{electromagnetic}
\acro{HRLLC}{s high-rate and high-reliability low latency communications}
\acro{6DoF}{six degrees of freedom}
\acro{MR}{mixed reality}
\acro{PAPR}{peak to average power ratio}
\acro{OFDM}{orthogonal frequency-division multiplexing}
\acro{OFDMA}{orthogonal frequency-division multiple access}
\acro{SC-FDM}{single carrier frequency-division multiplexing}
\acro{ToA}{time of arrival}
\acro{MUSIC}{multiple signal classification}
\acro{IoE}{Internet of Everything}
\acro{CRLB}{Cramér–Rao lower bound}
\acro{QoPE}{quality of personal experience}
\acro{AI}{artificial intelligence}